\documentclass[11pt, letterpaper]{article}

\usepackage[utf8]{inputenc}
\usepackage[margin=1.5cm]{geometry}
\usepackage{titlesec}
\usepackage{tabu}
\usepackage{enumitem}
\usepackage{amssymb}
\usepackage{xcolor}
\newlist{selectlist}{itemize}{2}
\setlist[selectlist]{label=$\square$,leftmargin=*,noitemsep,topsep=0pt}
\usepackage{graphicx}
\usepackage{longtable}
\usepackage[numbers,sort&compress]{natbib}
\usepackage{algorithm}
\usepackage{siunitx}
\usepackage{lmodern}
\usepackage{url}
\usepackage{mhchem}

\usepackage{hyperref}
\usepackage{xr-hyper}
\hypersetup{
    colorlinks=true,
    linkcolor=blue,
    filecolor=magenta,      
    urlcolor=blue,
}
\usepackage{xcolor} % 1. Required package for colors
\titleformat{\section}[block]{\hspace{1em}\bfseries}{\thesection.}{0.5em}{} 
\titleformat{\subsection}[block]{\hspace{1em}\bfseries}{\thesubsection}{0.5em}{}

\begin{document}
% \myexternaldocument{Supplementary}
% Create the title block
% \begin{flushleft}

% Remove all text in italics when filling out the template and replace with your manuscripts corresponding text in regular font.
%\textit{Text in italics are template instructions. Remove and replace all instructions with regular font text.}

\setlength{\parindent}{0pt}
\setlength{\parskip}{10pt}
% \textbf{\large HardwareX article template}

%Insert title
%Max. 20 words. A good title should contain the fewest possible words that adequately describe the content of a paper.
\newpage
\textbf{ARTGEL: A temperature-regulated electrophoresis platform for quantitative studies of reversible association in gels}\\ 
% \textbf{ARTGEL: An Open-Source Platform for Independent Thermal Control and Constant-Voltage Operation across gel in Long-Duration Agarose Gel Electrophoresis}\\ 
% \textit{Please avoid acronyms and abbreviations where possible.}
% ARTGEL: Automated Reguated Temperature Gel Electrophoresis Lab-system

%Insert Authors
\textbf{Authors}\\ \textit{Rupam Saha, Seth Fraden*}

%Insert Affiliations
\textbf{Affiliations}\\ \textit{Martin A. Fisher School of Physics, Brandeis University, Waltham, Massachusetts 02453, USA}

%Insert Contact Email
%Include institutional email address of the corresponding author
\textbf{Corresponding author’s email address}\\ \textit{fraden@brandeis.edu}

%Insert Abstract
%Max. 200 words. Remember that the abstract is what readers see first in electronic abstracting and indexing services. This is the advertisement of your article. Make it interesting, and easy to be understood. Be accurate and specific, keep it as brief as possible.
\textbf{Abstract}\\ 
Here we present ARTGEL, an actively regulated-temperature gel electrophoresis platform designed for long-duration experiments under independently controlled thermal and electrical conditions. ARTGEL combines thermoelectric regulation of the gel temperature, a large heated and circulated buffer reservoir, and an automated electrode-wiping mechanism that stabilizes the voltage across the gel during runs exceeding 24 h. The platform was developed to address a limitation of conventional electrophoretic mobility shift assays, which are commonly used to analyze reversible biomolecular association but usually aim to suppress reaction during electrophoresis by dilution, competitors, or reduced temperature so that the gel reports a pre-equilibrated bulk solution. For temperature-sensitive systems, these strategies can alter the chemical state during loading and migration and obscure whether the measured band pattern reflects the original bulk sample or a re-equilibrated state inside the porous gel. Rather than attempting to quench reactions, ARTGEL enables electrophoresis to be performed at the same temperature as complementary bulk measurements, so that reversible association can be quantified directly in the gel and compared with matched measurements in solution. Using DNA origami assemblies, we show that ARTGEL preserves distinct temperature-dependent association states, resolves reaction-dependent distortions of migrating bands, and supports extraction of in-gel kinetic and thermodynamic parameters from reaction–diffusion–advection modeling.

%Insert Keywords
% At least 3 keywords. There is no limit on the no. of keywords you can list. Please remember that effective keywords should not repeat words appearing in your title, and should be neither too general nor too narrow.

\section{Hardware in context}
% Include a short description of the hardware, putting into context of similar open hardware and proprietary equipment in the field.

Gel electrophoresis is one of the most widely used laboratory techniques for separating nucleic acids, protein-nucleic acid complexes, and nanoscale biomolecular assemblies based on electrophoretic mobility. In most laboratory settings, it is used as a rapid analytical method to report molecular size distributions or sample purity under convenient ambient conditions. By contrast, an electrophoretic mobility shift assay (EMSA) is used to analyze interacting species, often with the goal of extracting binding equilibria and, in some cases, kinetic information for reversible biomolecular complexes~\cite{fried1981equilibria, fried1989measurement, gerstle1993measurement, fried1998electrophoretic,  schuck2007protein, hellman2007electrophoretic, shubsda1999characterization, shubsda2000binding, jarmoskaite2020measure}. 

In its conventional form, EMSA is usually intended to report the equilibrium state established in bulk solution before loading. The common strategy is therefore to make the gel as kinetically silent as possible: association is suppressed by dilution or competitors~\cite{straney1987kinetics, recht2001central, ross2009analysis}, and dissociation is reduced by lowering temperature or by chemical stabilization~\cite{gerling2018sequence}. This strategy assumes that the population distribution prepared before loading remains effectively unchanged during sample loading, entry into the gel, and subsequent migration.

In practice, that assumption is often difficult to justify. Sample loading and entry into the gel can require 10-20 min (Fig.~\ref{sfig:pokt_exit}), which may be sufficient for substantial re-equilibration, especially when the gel is operated at a temperature different from the incubation condition. Lowering concentration to suppress association will reduce the signal-to-noise ratio and lowering temperature can increase  association. More generally, even when electrophoresis is performed carefully, it may be unclear whether the observed band pattern reports the pre-equilibrated bulk solution or a state that has evolved further inside the porous gel matrix.

Here, we are motivated by a different scientific question. Rather than attempting to make the gel kinetically silent, we seek to make it a well-defined and controllable reaction environment whose behavior can be compared directly with bulk solution. In this framework, the gel is not treated as an inert readout of a pre-equilibrated sample: confinement may alter $k_{\mathrm{on}}$, $k_{\mathrm{off}}$, and, in principle, $K_d$. ARTGEL therefore enables two complementary measurements. First, because electrophoresis spatially separates monomers, dimers, fragments, aggregates, and higher-order assemblies, it directly tests whether the assumed association model is adequate. Second, by operating at the same temperature as matched bulk measurements, ARTGEL allows the in-gel equilibrium and rate constants to be measured and compared with their solution values. For the DNA origami~\cite{rothemund2006folding} system studied here, this comparison shows that the bulk and in-gel values of $K_d$ are similar, whereas the association and dissociation rates differ by more than an order of magnitude, demonstrating that the gel primarily alters the kinetic pathway rather than the equilibrium association state.

To pursue this approach quantitatively, electrophoresis must be performed under conditions where temperature, electric field, and buffer composition remain well controlled over long durations. Under such conditions, the evolution of reactive species inside the gel can be analyzed by fitting spatial intensity profiles measured as a function of time to reaction-diffusion-advection models~\cite{shubsda1999characterization, shunong1991gel, kleinschmidt1991computer, cann1996theory, cann1998theoretical}. This requires, at minimum, (i) constant and spatially uniform temperature, because reaction rates are highly temperature sensitive, and (ii) a stable electric field, because drift velocity enters directly into the transport model. For long experiments in high-salt buffers, maintaining approximately constant ionic conditions is also important for reproducible migration, thermal control and maintaining constant interparticle interaction strength.

Conventional electrophoresis platforms are not designed to meet these requirements. Joule heating in buffers containing added salt, including MgCl$_2$-containing buffers used for DNA origami, raises the gel temperature, broadens bands, and perturbs both mobility and reaction kinetics~\cite{hellman2007electrophoretic, slater1995diffusion, sajjadi2019heat}. Attempts to regulate temperature by adjusting electrophoresis power inherently couple thermal control to electric field strength, preventing independent control of these variables~\cite{labsmith_uEP01}. In addition, prolonged electrophoresis in high-salt environments promotes electrode fouling through electrochemical precipitation, for example Mg(OH)$_2$ in the presence of MgCl$_2$, which increases electrode resistance, changes the effective voltage across the gel (SI section~\ref{sec:tworesistormodel}), and alters buffer composition over time. These effects compromise quantitative reproducibility during long-duration runs.

To address these limitations, we developed ARTGEL, an \textbf{a}ctively \textbf{r}egulated-\textbf{t}emperature  \textbf{g}el \textbf{el}ectro-phoresis platform for long-duration experiments that provides independent control of gel temperature and electric field. ARTGEL integrates (i) direct thermoelectric regulation of gel temperature using a Peltier module coupled to an aluminum heat-spreading plate under proportional-integral (PI) control, (ii) an actively heated and continuously circulated large volume of buffer to reduce the thermal load on the Peltier module, improve spatial temperature uniformity, and minimize the decrease in ionic strength (SI section~\ref{sec:salt_depletion}), and (iii) an automated electrode-wiping mechanism that mitigates salt deposition and stabilizes electrode resistance. Together, these features create a controlled in-gel environment suitable for extracting in-gel kinetic and thermodynamic parameters and for comparing them directly with complementary bulk measurements~\cite{wei2024hierarchical}.

ARTGEL is modular, fully documented and constructed from widely available components. All hardware designs, firmware, and control software are openly shared, enabling straightforward reproduction and customization. By transforming gel electrophoresis from a primarily qualitative separation method into a controlled platform for reactive transport in porous media, ARTGEL expands the utility of gel-based assays for quantitative studies of reversible biomolecular association, confinement effects, and temperature-sensitive nanoscale assemblies.

\begin{figure*}[th!]
 \centering
 \includegraphics[width=1.0\linewidth]{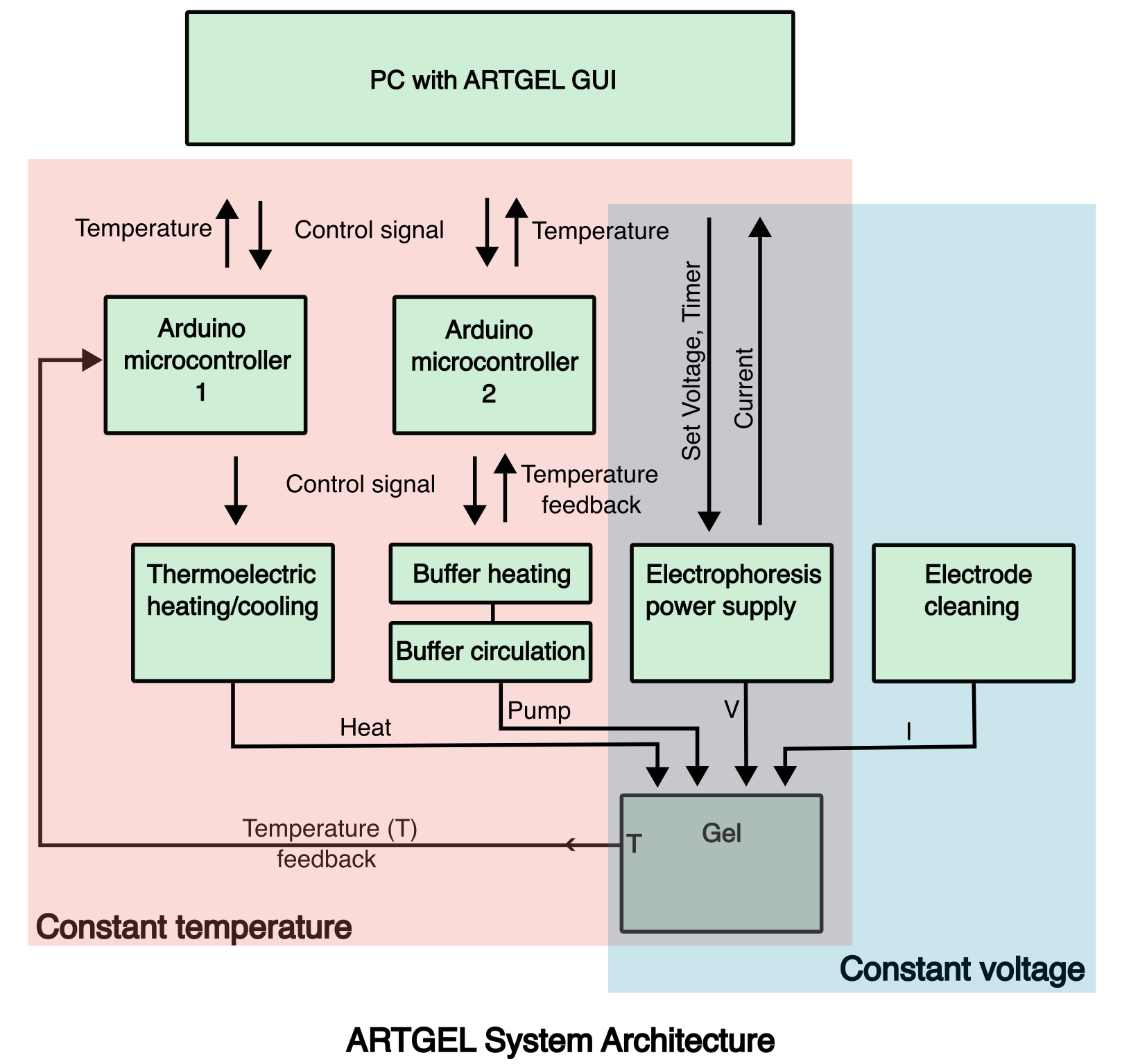}
 \caption{\textbf{Schematic of the gel temperature control  architecture.} Red denotes parts of the system that contribute to controlling the gel temperature through the flow of heat. Blue denotes system elements that control the gel voltage. The gel receives four physical perturbations from the ARTGEL that affect its temperature; two are heat from the thermoelectric and circulating buffer, two set the electrophoretic electrical dissipation, $VI$, in which the electrophoresis power supply sets the voltage, $V$, and electrode cleaning maintains constant current, $I$. Passive heat exchange with the environment is not indicated.} 
 \label{fig:architecture}
\end{figure*}

\begin{figure*}[th!]
 \centering
 \includegraphics[width=1.0\linewidth]{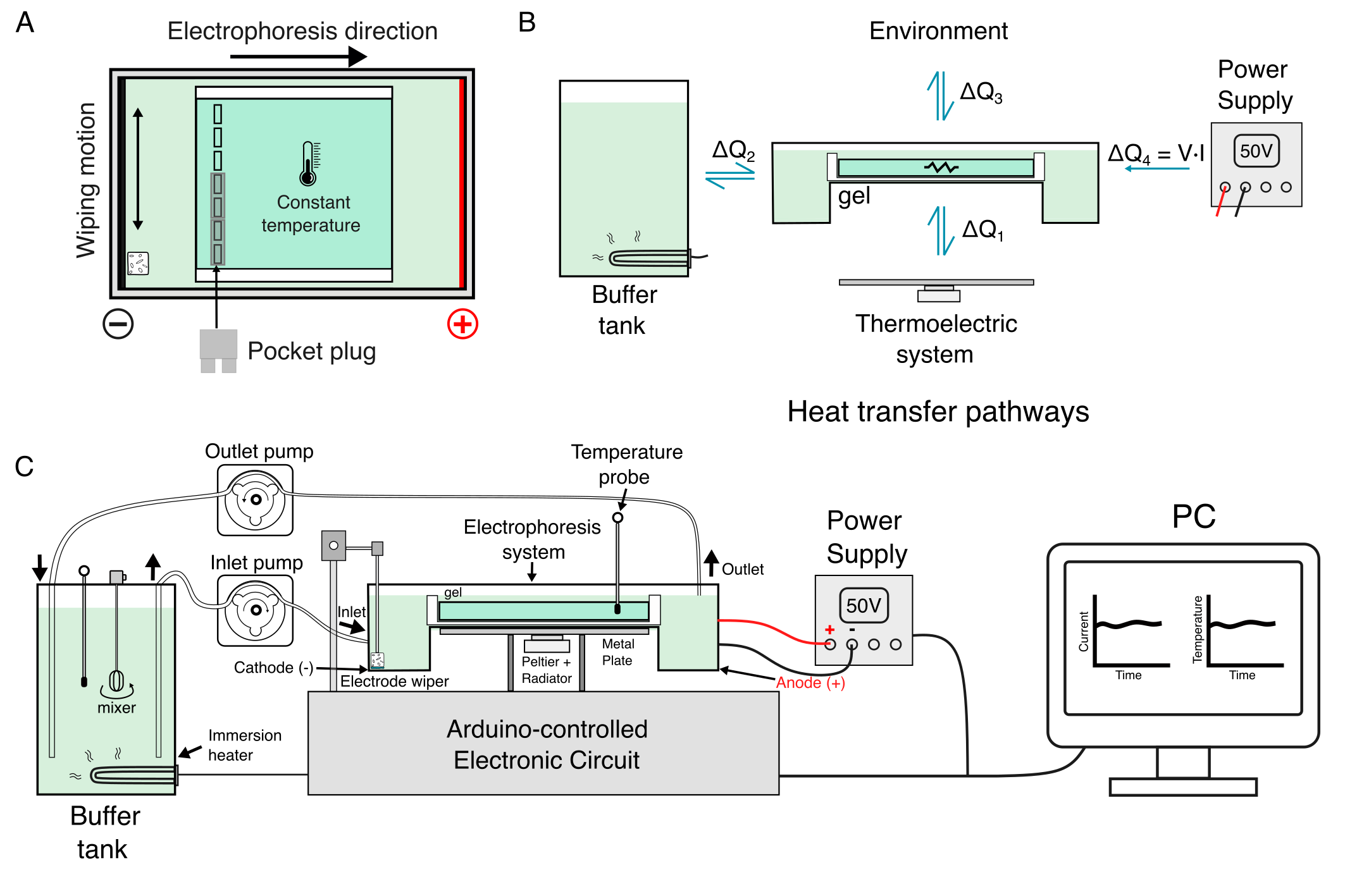}
 \caption{\textbf{System layout} %\sfred{Question: Why did you place the outlet of the buffer on the cathode side, rather than the anode side? If you placed the outlet on the anode side, would that prevent the salt precipitate from entering the pockets?}
 A. Top-down view of the gel during electrophoresis under constant temperature and constant voltage conditions. B. Schematic of heat-transfer pathways between the gel and surrounding components and electrical heat generation, including the buffer tank, thermoelectric system, external environment and electrophoresis power supply. 
 C. ARTGEL schematic.
 }
 \label{fig:summary}
\end{figure*}

\begin{figure*}[th!]
 \centering
 \includegraphics[width=1.0\linewidth]{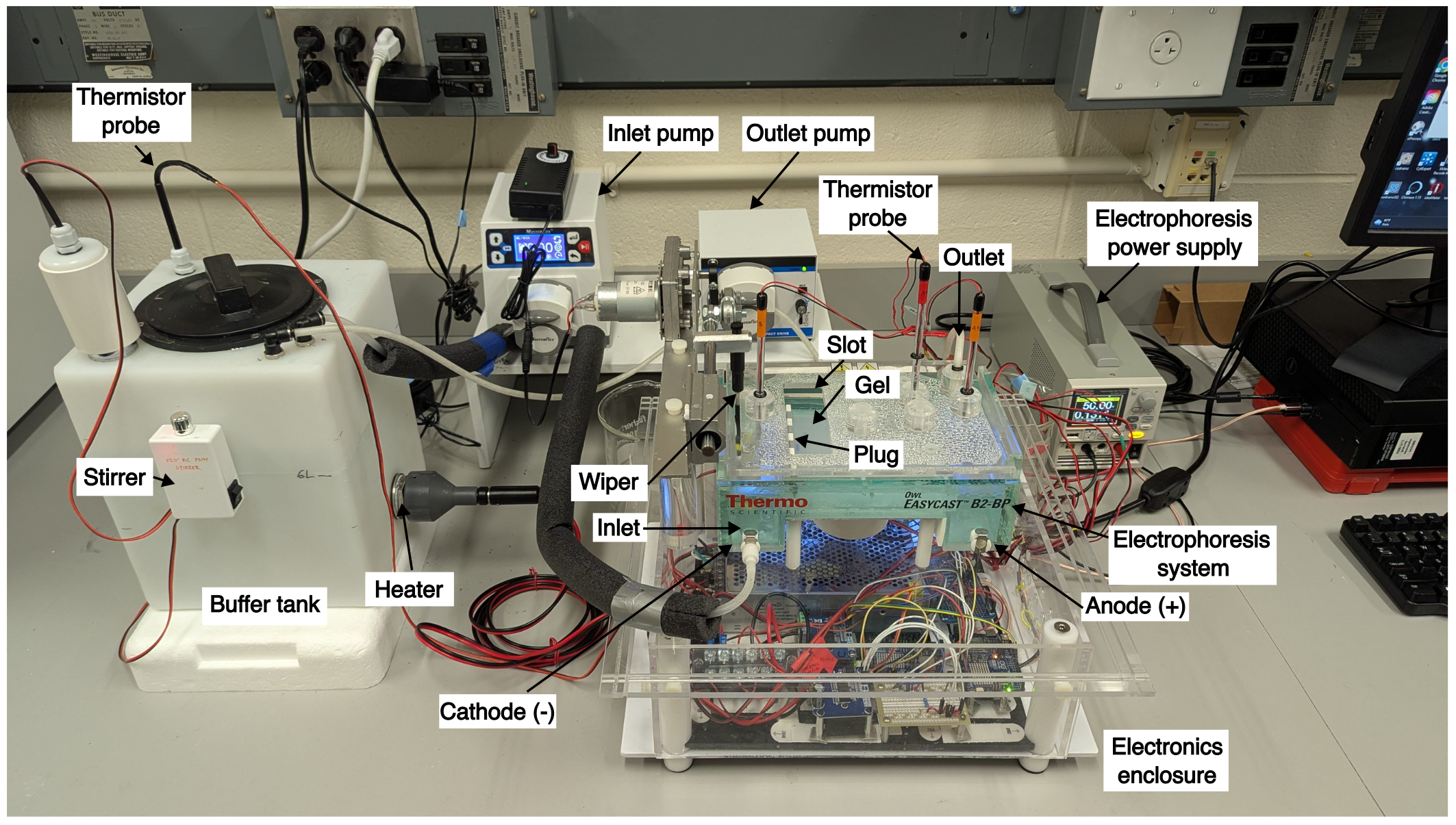}
 \caption{\textbf{Instrument} 
 }
 \label{fig:instrument}
\end{figure*}

\section{Hardware description}
% Describe the hardware, highlighting the customization rather than the steps of the procedure. Highlight how it differs/which advantage it offers over pre-existing methods. For example, how could this hardware: be compared to other hardware in terms of cost or ease of use, be used in the development of further designs in a particular area, and so on.

% > Add 3-5 bulleted points to broadly explain to other researchers how the hardware could be potentially useful to them, for either standard or novel laboratory tasks, inside or outside of the original user community.

\subsection{Hardware design and construction}

In ARTGEL during electrophoresis we consider the gel to thermally interact with four components; the circulating buffer flowing over the gel, the thermoelectric (Peltier) module mounted outside and just below the gel, the electrophoresis power supply, which generates a voltage drop across the gel and concomitantly, resistive heating of the gel, and the ambient lab environment to which heat from the gel leaks (Fig.~\ref{fig:summary}B). In our applications, the gel is typically operated about 10 $^{\circ}$C above room temperature. In this circumstance, in order of thermal mass, the gel is primarily heated by the circulating warm buffer and next the electrophoresis power supply. For active temperature control we employ a thermoelectric module under the gel which can actively supply or remove  heat from the gel to regulate the gel temperature at the desired set point.
Constant voltage operation across the gel is maintained by periodically removing salt deposits from the negative electrode by an automated wiping mechanism, described in more detail in Sec. \ref{electrodewiper}. A  supplementary movie shows the instrument in operation (SI section~\ref{SI:movie}). 

% ARTGEL is a low cost, automated and regulated temperature-controlled gel electrophoresis device. This device enables users to run agarose gel at constant temperature and constant current for a sufficiently longer duration. For uninterrupted gel electrophoresis for an extended duration, we use a 2 Gallon buffer tank fitted with an immersion heater at the bottom. while the gel recieves heat from the warm buffer, for fine tuning of gel temperature, we implement Peltier element. Both immersion heater and the peltier are controlled by microcontroller which is Arduino in this case. To maintain constant current, we employ a linear actuator which periodically translates a wiper to clean the deposited salt from the negative electrode. Thermistors are used as temperature probes to measure the temperature of the gel, buffer and room. We incorporate MATLAB for data logging and to control the temperature and circuit current in real time. 

\subsubsection{Mechanical architecture}

\begin{figure*}[th!]
 \centering
 \includegraphics[width=1.0\linewidth]{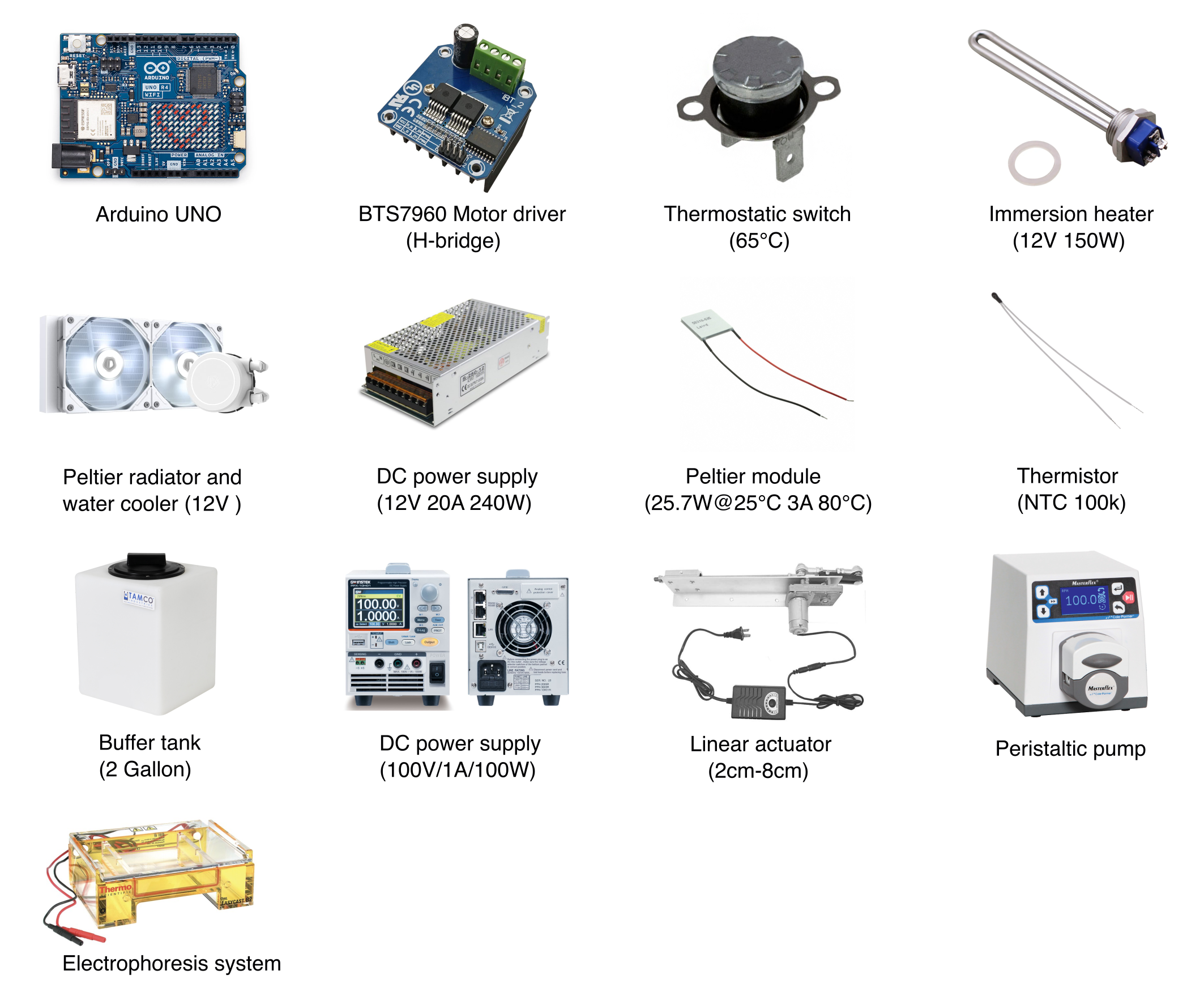}
 \caption{\textbf{Commercial ARTGEL components.} 
 }
 \label{fig:apparatus}
\end{figure*}

The electrophoresis part of the ARTGEL system is physically organized as a stacked modular assembly (Fig.~\ref{fig:summary}C). The gel electrophoresis box is mounted on top of a water-resistant electronics enclosure (Figs.~\ref{fig:instrument} and~\ref{fig:machined_apparattus}B).  The enclosure is constructed using Plexiglas sheets  and nylon standoffs and houses all control electronics including microcontrollers, motor drivers (H-bridge), breadboards and power supply (Fig.~\ref{fig:apparatus}). Ventilation openings on the enclosure walls allow airflow for thermal dissipation.

The top surface of the enclosure extends beyond the side walls and incorporates integrated drainage channels like moats on both the upper and lower faces~(\ref{fig:machined_apparattus}B). These channels guide accidental liquid spills toward a corner reservoir, preventing fluid ingress into the electronics.

 A flat aluminum heat-spreading plate is attached to the bottom of the gel electrophoresis box and is mounted on four standoffs on top of the electronics enclosure. A thermoelectric module and radiator assembly are attached to the underside of this plate. The area of the plate matches the area of the gel box to ensure spatially uniform heat transfer (Fig.~\ref{fig:summary}C). 

\subsubsection{Thermal regulation subsystem}
Thermal regulation in ARTGEL is implemented using a two-stage approach. Coarse temperature control is achieved by circulating 8L of temperature-controlled buffer from an external reservoir through the gel box. The buffer tank contains a 150W immersion heater. Typically we set the buffer tank at a temperature, 0 - 2$^{\circ}$C higher than the set point temperature of the gel. Heat is also supplied by electrophoresis itself. We typically use 50V and with 20 mM MgCl$_2$ buffer, generating a current of about 140 mA, providing  7 W of heat to the gel. Fine temperature adjustment is provided by a 24~W thermoelectric (Peltier) module coupled to the aluminum plate beneath the gel, which is used to heat or cool the gel to the desired set point.

Prior approaches to temperature regulation in gel electrophoresis have relied on circulating heated buffer to regulate gel temperature~\cite{hansen1992electrophoresis, tippins1999apparatus}. While effective in delivering heat, large buffer volumes introduce substantial thermal inertia, leading to slow response times and limited tunability. Conversely, thermoelectric control alone offers rapid adjustment but lacks the thermal uniformity provided by bulk buffer. In ARTGEL, these components work synergistically: the buffer reservoir provides thermal stability and spatial uniformity, while the thermoelectric module enables rapid and precise temperature tuning.

Temperature is monitored using multiple 100k\si{\ohm} thermistors positioned at key locations, including within the gel, in the aluminum plate, in the buffer reservoir, and in the ambient environment. Each thermistor is entirely electrically insulated. Those embedded in the gel are housed in a cylindrical plastic enclosure with a nickel tip to provide good thermal contact with the gel, as well as corrosion resistance~(Fig. \ref{fig:machined_apparattus}A).

The thermoelectric module can be controlled manually, or automatically with proportional or proportional–integral (PI) feedback implemented on a microcontroller, referred to as Arduino~1. The buffer tank is controlled by a separate PI controller referred to as Arduino~2 (see SI section~\ref{sec:PItuning}). A thermal safety switch rated at $65^{\circ}$C is attached to the thermoelectric heat spreading plate to prevent overheating from user or software errors.

\subsubsection{Electrophoresis and electrode maintenance subsystem}
\label{electrodewiper}
Electrophoresis is performed in a standard Thermo Scientific Owl Easycast B2-BP horizontal agarose DNA device. The device has two electrodes, each located in a deep buffer-filled well. The agarose gel, 1 cm thick, separates the two electrodes. One can model the effective resistance of the material between the electrodes in a lumped circuit model consisting of three resistances; the resistance between the anode and the gel, the resistance across the gel and the resistance between the gel and the cathode. We want the voltage drop across the gel to be constant so that the molecules migrate at constant velocity. However, the constant voltage power supply acts between the electrodes instead of across the gel. Consequently, if either of the electrode resistances changes with time, then the voltage drop across the gel will also change in time, even if the voltage between the electrodes remains constant. Because ARTGEL is an externally driven electrolytic cell, the cathode is the negative electrode and the anode is the positive electrode. Our buffer contains MgCl$_2$, which leads to the electrochemical deposition of a thick, white coating of  Mg(OH)$_2$ on the cathode, which increases electrode resistance, lowering the current and hence the voltage drop across the gel. The ARTGEL supports constant-voltage electrophoresis across the gel over extended durations by removing accumulated salt  at the negative electrode using a custom-built reciprocating linear actuator  (Fig.~\ref{fig:instrument} and \ref{fig:machined_apparattus}F).

The actuator drives a scrubbing sponge attached to the wiper that periodically translates along the length of the negative electrode. The stroke length is adjustable, allowing the sponge to contact the full electrode length. The wiping frequency is set to approximately one stroke every five to ten seconds, which is found to effectively prevent salt buildup during long runs. The stroking speed is controlled using an integrated potentiometer.

We also designed and 3D printed plugs that fit into the wells of the gel. The purpose of the plugs is to prevent the scraped solid salt particles from the negative electrode from entering the gel. See SI section~\ref{sec:plug} for more details.

A commercial programmable electrophoresis power supply is used to operate the gel under constant voltage, and the voltage and the current are logged in real time.

\subsubsection{Buffer circulation and reservoir}
Continuous buffer circulation is provided using an external 8L buffer reservoir connected to the gel box via tubing and two peristaltic pumps (Fig.~\ref{fig:summary}C and \ref{fig:instrument}). Buffer enters the gel box through an inlet port on the side wall and exits through outlet tubing inserted into an outlet port integrated into the lid (Fig. \ref{fig:summary}C). The height of the outlet tubing opening sets the height of the buffer in the gel box. The inlet pump is run at 30~ml/min. The outlet pump is operated at a higher flow rate than the inlet pump, ensuring that the buffer height within the gel box remains regulated and preventing overflow. Inside the reservoir, a 150 W electrical immersion heater provides bulk heating, while a DC motor-driven stirrer maintains uniform buffer temperature (\ref{fig:machined_apparattus}C). Cooling is passive through heat loss to the environment. The buffer temperature is measured using an independent thermistor and regulated using a dedicated microcontroller, referred to as Arduino~2.

% \subsubsection{Buffer tank}
% We customize the buffer tank to ensure warm buffer is circulated in gel box for longer duration. The buffer tank contains a pair of inlet and outlet which are connected with the gel box via tubings. At the bottom of the tank, an electrical immersion heater is fitted to heat the buffer. Additionally, the tank contains a 12V DC motor to stir the buffer. Finally, we incorporate a temperature probe to measure the buffer temperature. Both the immersion heater and probes are controlled using arduino placed inside the enclosure.

\subsubsection{Electronics and control hardware}
ARTGEL uses two Arduino-based microcontrollers operating independently, referred to as Arduino 1 and 2. Arduino~1 regulates gel temperature via the thermoelectric module, while Arduino~2 regulates the buffer temperature in the reservoir. Each Arduino interfaces with a motor driver (H-bridge) that serves to switch the voltage direction and through PWM, set the average voltage. The two H-bridges are powered by a shared 12 V, 250 W power supply, referred to as power supply 1. The Arduino uses pulse width modulation (PWM) to control the voltage to the thermoelectric. The PWM frequency is 976 Hz, which is much higher than the thermal response rate of the gel box, so the temperature of the gel is set by the PWM value and not the instantaneous power.

Power supply 1 also drives the cooling fan and radiator assembly associated with the thermoelectric module. Five thermistors (A0$^1$, A2$^1$, A3$^1$, A4$^1$ and A5$^1$) are connected to the gel temperature controller, while one thermistor is connected to the buffer temperature controller.

\begin{figure*}[th!]
 \centering
 \includegraphics[width=0.9\linewidth]{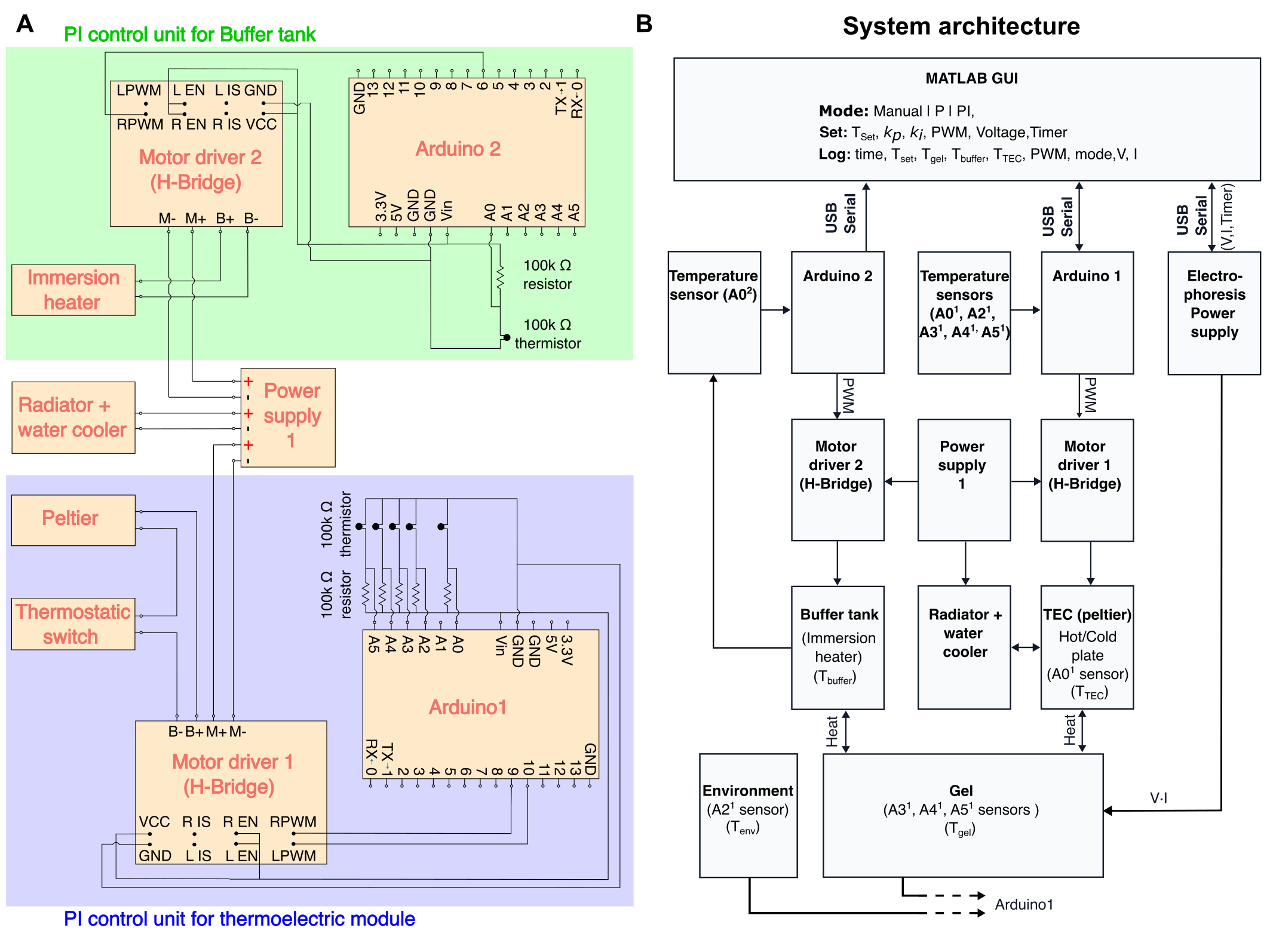}
 \caption{\textbf{Pin configuration and system architecture.} 
 (A) Circuit diagram illustrating the Arduino pin configuration used to interface with the various electronic components.
(B)  Block diagram depicting the overall system architecture, showing how individual components interact with one another and are controlled via MATLAB.  }
 \label{fig:circuit}
\end{figure*}

A computer interface enables real-time data logging and control via MATLAB. Custom software provides live visualization of temperature, voltage, and current data, as well as manual,  closed-loop proportional and proportional-integral temperature control. See Fig.~\ref{fig:circuit} for the pin configuration and system architecture. 

% The circuit is driven by two arduino microcontrollers. While one arduino is used to regulate the gel temperature, the other one is used to regulate the buffer temperature. Each arduino is connected to a Hbridge relay which is connected to a 12V, 250Watt power supply. The PCB design can be found ... To regulate the gel temperature we incorporate a 24Watt peltier and a 60C thermal switch in the circuit. We include 5 100K thermistors to in this Arduino. While 4 of the thermistors are used to measure the gel and room temperature, the 5th thermistor is utilized to measure the metal plate temperature. The second arduino is integrated with 150W immersion heater and one thermistor to regulate the buffer temperature in the tank. The power supply also supplies power to the cooling system consisiting of the fan and the radiator. 
% For electrophoresis we use Gwintek 100V power supply. This power supply can be programmable to both constant voltage and current as required. We integrate it with PC to record the voltage and current in the circuit.

\subsection{Graphical user interface}
\begin{figure*}[th!]
 \centering
 \includegraphics[width=0.9\linewidth]{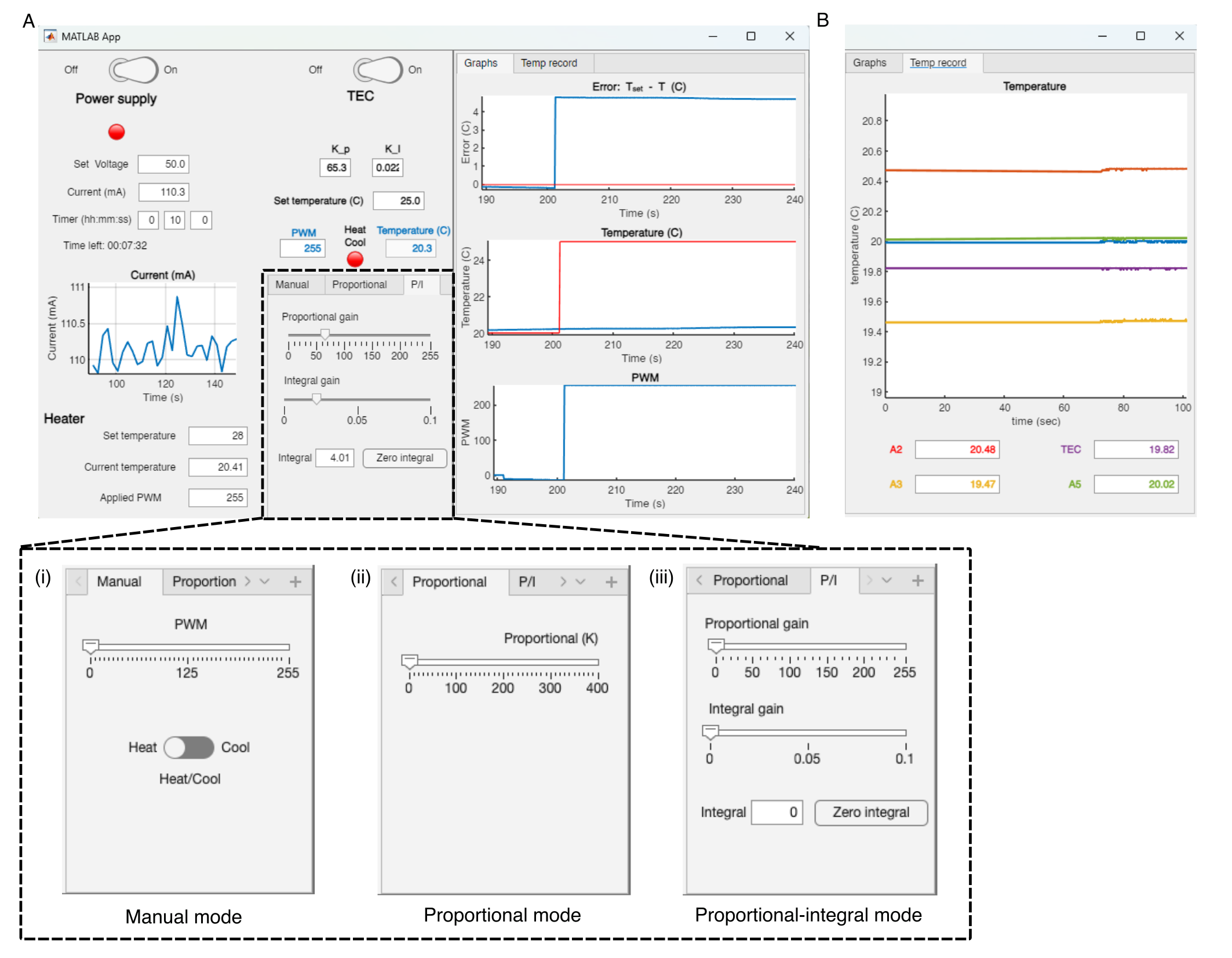}
 \caption{\textbf{Graphical User Interface}
 }
 \label{fig:GUI}
\end{figure*}

The graphical user interface (GUI) is developed in MATLAB to provide centralized control, real-time monitoring, and automated data logging for the electrophoresis instrument (Fig.~\ref{fig:GUI}). Through the GUI, users can specify key experimental parameters, including target gel temperature, control mode (manual, P, PI), electrophoresis voltage, and run duration. These inputs are transmitted to an Arduino-based controller via serial communication. During operation, the GUI continuously displays real-time measurements of electrophoresis current, gel temperature, and buffer temperature, together with the corresponding pulse-width modulation (PWM; 0–255) signals used for active temperature regulation.

Both open-loop and closed-loop temperature control strategies are implemented within the system. In manual control mode, the PWM value is directly set by the user, providing open-loop control of heating or cooling (Fig.~\ref{fig:GUI}A(i)). In PI control mode, the PWM output is automatically adjusted in a closed-loop to maintain the gel temperature at the desired setpoint Fig.~\ref{fig:GUI}A(ii). Live plots update dynamically throughout the experiment, enabling the user to assess temperature stability, transient responses, and controller performance in real time (Fig.~\ref{fig:GUI}B). All experimental parameters—including voltage, current, temperature, PWM, and time—are automatically recorded into structured data files for post-experiment analysis. In addition, integrated start and stop controls allow safe initiation and immediate termination of experiments. Together, the GUI provides an intuitive and robust interface that seamlessly integrates user input, hardware control, and quantitative data acquisition for temperature-regulated agarose gel electrophoresis.

\section{Build instructions}
%%%%%%%%%%%%%%%%%%%%%%%%%%%%%%%%%%
%Provide detailed, step by step instructions for the construction of the reported hardware include all necessary information for reproducing the submitted hardware.
% > Explain and, when possible, characterize design decisions. Including design alternatives if they exist. 
% > Use visual instructions such as schematics, images, and videos. 
% > Clearly reference design files and component parts described in the Design File Summary and Bill of Materials. 
% >Highlight potential safety concerns that may arise
%%%%%%%%%%%%%%%%%%%%%%%%%%%%%%%%%%

\begin{figure*}[th!]
 \centering
 \includegraphics[width=1.0\linewidth]{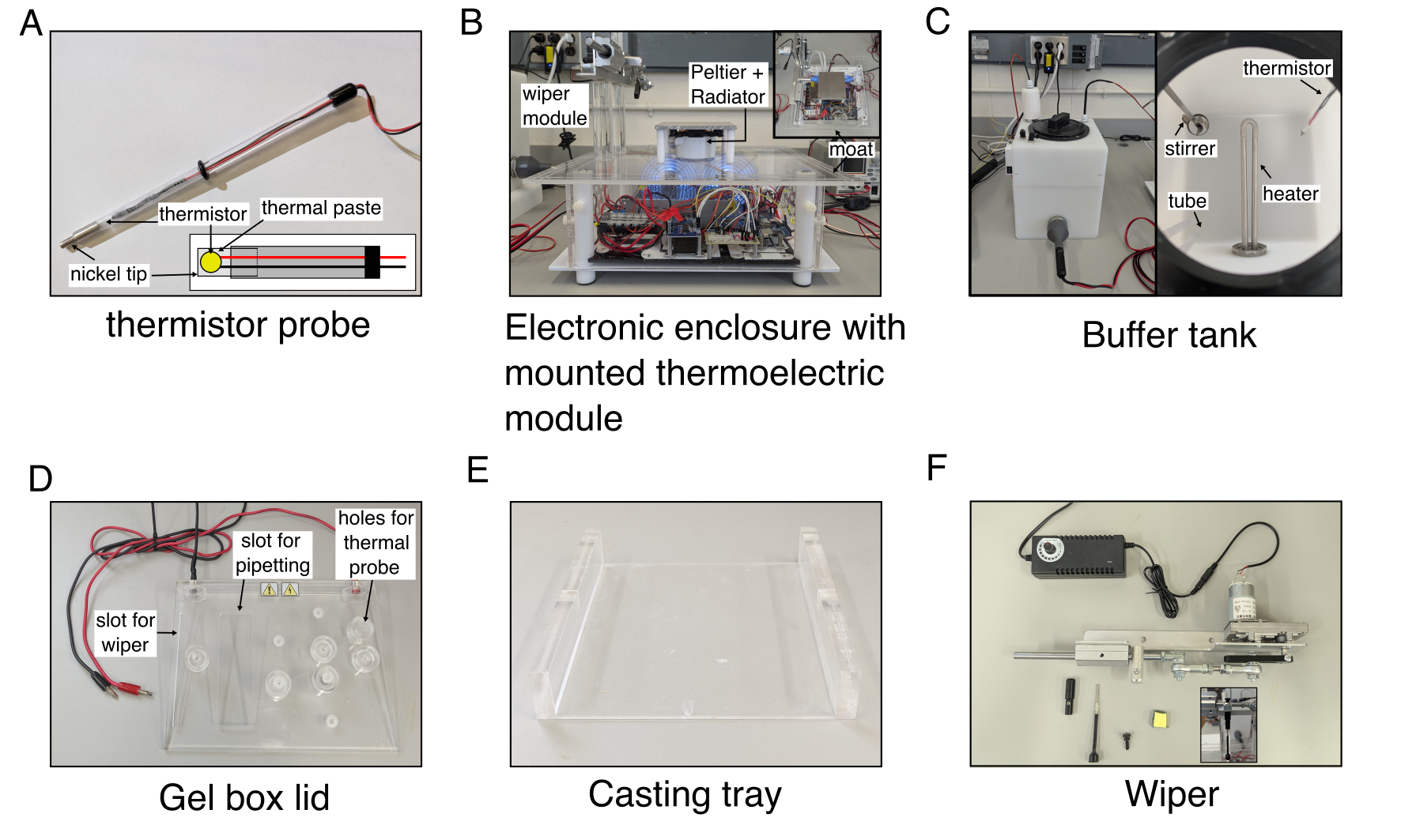}
 \caption{\textbf{Customized ARTGEL components.}
 }
 \label{fig:machined_apparattus}
\end{figure*}

1. To enable temperature monitoring during electrophoresis, six custom-fabricated thermistor probes are designed with appropriate electrical insulation and mechanical protection for operation in conductive and aqueous environments. The six probes are positioned to monitor the buffer reservoir (A0$^2$), ambient room temperature (A2$^1$), the aluminum plate (A0$^1$), and three distinct locations within the gel for direct temperature measurement (A3$^1$, A4$^1$ and A5$^1$). Of these, four probes (the three gel probes and the buffer reservoir probe) require electrical insulation due to direct contact with conductive or aqueous environments (Fig.~\ref{fig:machined_apparattus}A). Each thermistor is first soldered to 26-AWG insulated wires, and the solder joints are mechanically reinforced and electrically insulated using heat-shrink tubing. The wired thermistor is then inserted into a hollow plastic tube (approximately 6~mm  in external diameter and 16 cm in length), with the thermistor leads exiting from the opposite end of the tube. A nickel tip, machined with an outer diameter slightly smaller than the inner diameter of the plastic tube, is used to house the exposed thermistor bead.  Thermal adhesive is applied to the thermistor tip, which is then sealed inside the nickel tip to ensure good thermal contact while maintaining electrical isolation. The nickel tip is subsequently inserted or snap-fitted into the plastic tube. The open end of the tube is sealed using a rubber cap, which both secures the wiring and reduces mechanical stress on the leads. For the buffer reservoir probe, the plastic tube length was increased as required to accommodate the reservoir geometry and immersion depth.

Precaution: Make sure the nickel tip is tightly attached and sealed to the tube to prevent the water leakage. One can use vacuum grease as sealant. 

2. The electrical circuit inside the electronics enclosure consists of two Arduino microcontrollers, two motor driver (H-bridge) modules, two breadboards, six 100k\si{\ohm} resistors, six 100k\si{\ohm} thermistors, one terminal block, and a DC power supply (referred to as ``Power supply 1").  The motor driver functions as a switch to reverse the direction of current, enabling the thermoelectric module to both heat and cool the gel. One Arduino–motor driver (H-bridge) pair is dedicated to controlling the thermoelectric (Peltier) module for gel temperature regulation, while the second Arduino–motor driver pair independently controls the immersion heater used for buffer temperature regulation.  Of the six 100~k\si{\ohm} resistors and six 100~k\si{\ohm} thermistors, five resistor–thermistor pairs are connected to Arduino 1, while one pair is connected to Arduino 2. Each known resistor forms a voltage divider with its thermistor, allowing the Arduino to determine temperature from the measured voltage.  See SI section~\ref{SI section: calibration} for a detailed description on thermistor calibration. Two solderable breadboards are used to keep the two control circuits electrically separated. All the components are assembled following the wiring schematic shown in Fig.~\ref{fig:circuit}A, with all control and power connections verified prior to integration. Power supply 1 is connected to a 10~A, 125~V power cord cable and provides power to the cooling fan–radiator assembly, the Peltier module via its motor driver, and the immersion heater via its corresponding motor driver. A terminal block is used to organize signal wiring, reducing clutter and simplifying troubleshooting. All electronic components are mounted on a rectangular plastic platform using standoffs and secured with hook-and-loop fasteners to ensure mechanical stability while allowing easy access for maintenance and modification.

3. The electronics enclosure is constructed using a stacked Plexiglas architecture supported by four machined nylon standoff pillars, each approximately 12~cm in height (Fig.~\ref{fig:machined_apparattus}B). The bottom rectangular base plate has dimensions of 40~cm × 30~cm, while the top plate is intentionally fabricated with a larger footprint and curved edges incorporating integrated drainage channels. These channels guide accidental liquid spills away from the electronics and toward an external collection reservoir. Each standoff pillar contains curved internal slots that allow the electronics mounting platform to slide into the enclosure during assembly. The left, right, and front enclosure walls are fabricated from perforated Plexiglas sheet to promote airflow, whereas the rear wall houses the cooling fans. The top plate is secured to the four main standoffs using screws. Four additional short standoffs (approximately 4 cm in height) are then mounted on the top surface of the enclosure to support the aluminum thermal plate. The aluminum plate is secured to these standoffs such that it sits above the enclosure roof, forming the thermal interface for the gel box. The thermoelectric (Peltier) module, thermal safety switch, and radiator are mounted to the underside of the aluminum plate. A hole is drilled into the aluminum plate from the edge to the center to house the thermistor (A0$^1$) used to monitor the plate temperature.

4. An 8 L (2-gallon) tank is used as the buffer reservoir (Fig.~\ref{fig:machined_apparattus}C). The immersion heater is mounted on one side wall of the tank approximately 3~cm above the bottom surface using a 1-inch NPSM (National Pipe Straight Mechanical) threaded flange. The heater is configured with a female threaded interface, and a custom male plug connected to 14-AWG insulated wire is fabricated to supply power while ensuring a secure and serviceable electrical connection. On the top surface of the reservoir, four access ports are drilled: one port accommodates the thermistor (A0$^2$) for measuring the buffer temperature, one is used to mount a 12~V DC mechanical stirrer for continuous mixing, and two additional ports serve as the inlet and outlet for buffer circulation tubing connected to the gel box. This configuration ensures uniform buffer temperature, reliable circulation, and ease of maintenance during long-duration electrophoresis experiments.

5. The gel box lid and gel tray are modified to accommodate temperature probes, the electrode wiper, and efficient thermal coupling between the gel and the aluminum thermal plate (Fig.~\ref{fig:machined_apparattus}D and E). First, multiple holes are drilled into the gel box lid directly above the gel footprint and above the two buffer reservoirs, allowing thermistors to be inserted vertically into the gel and buffer from above without disturbing the electrophoresis setup. Near the negative electrode, a rectangular slot approximately 40~mm wide and 115~mm long is cut into the lid to allow passage of the stick holding the wiper, enabling continuous cleaning along the full length of the electrode during operation. The gel tray is then machined to reduce its thickness to approximately 1~mm, minimizing thermal resistance and improving heat transfer from the underlying metal plate to the gel. Finally, all air gaps are sealed using compressible foam boards to provide thermal insulation and maintain stable electrophoresis conditions during extended runs.

6. The reciprocating linear actuator is customized to function as an automated electrode wiper (Fig.~\ref{fig:machined_apparattus}F). Two rigid standoffs are fabricated and mounted on the top surface of the electronics enclosure adjacent to the negative electrode side of the gel box. The linear actuator is securely screwed to these standoffs to ensure stable and repeatable motion during operation. A stainless-steel rod (length 10~cm) with a thicker round base is fabricated such that one end threads directly into the actuator arm, while the opposite end extends through the previously machined slot in the gel box lid (Fig.~\ref{fig:machined_apparattus}D). At the bottom of this rod, we tape a velcro hook that applies gentle pressure to a sponge element (Scotch-brite block of dimension 15~mm × 15~mm), transmitting the actuator’s motion. The rough surface of the sponge is oriented toward the negative electrode, where it gently removes salt deposits that accumulate during electrophoresis. The stainless-steel rod is covered with heat-shrink tubing to protect against corrosion from the electrolyte and to provide electrical insulation. The actuator is powered at 24~V and includes an integrated potentiometer, which allows adjustment of the motor speed and consequently the wiping stroke frequency.

7. Two peristaltic pumps are installed to establish continuous buffer circulation between the external reservoir and the gel box (Fig.~\ref{fig:summary}C). The inlet line from the buffer tank is routed through one peristaltic pump and delivers heated buffer into the gel box through a side port. A second line (outlet) from the gel box is routed through another peristaltic pump and the buffer is delivered back to the buffer tank. The outlet line is designed to exit the gel box vertically through the lid. The height of the opening in the outlet tube sets the height of the buffer to be 1-2 mm above the surface of the gel (Fig. \ref{fig:summary}C). The inlet flow rate (30~mL/min) is regulated to provide steady buffer delivery, while the outlet pump is intentionally operated at a higher flow rate to maintain a constant buffer height within the gel box. To further suppress fluctuations in liquid level, the outlet tubing is coupled to a 10~µL pipette tip, which acts as a flow restrictor. The inlet tubing is wrapped with insulating foam to minimize thermal losses during the transfer of heated buffer from the reservoir to the gel box.

8. Finally, the microcontrollers and the electrophoresis power supply are linked to  the computer using USB cables to actively monitor and regulate the temperature, heating, voltage and current. 

\section{Operation instructions}

\subsection{Buffer preparation and thermal control initialization}
\begin{itemize}
    \item Prepare the electrophoresis buffer according to the planned duration of the experiment. 

    \item Fill the buffer tank and insert the thermistor probe into the reservoir.

    \item Connect and plug the heater. 

    \item Turn on power supply 1 inside the electronics enclosure.

    \item Connect Arduino 2 to the desktop computer.

    \item Open Buffer\_temp\_control.ino program and:
        \begin{itemize}
            \item Set the target temperature
            \item Define the PI parameters (Kp, Ki)
            \item Compile and upload the code
        \end{itemize}

    \item Turn on the mechanical mixer to ensure uniform buffer temperature. 

    \item Open the serial monitor and verify that the buffer temperature reaches and stabilizes at the target value. 
\end{itemize}

\subsection{Gel Preparation and Thermal Coupling}
\begin{itemize}
    \item Prepare the gel following the standard protocol and allow it to fully set.
    
    \item Once the buffer temperature has reached and stabilized at the set value:
        \begin{itemize}
            \item Place the gel on the machined casting tray
            \item Position the gel box on the metal thermal plate
        \end{itemize}
        
    \item Apply a thin layer of thermal paste between the gel box and metal plate to improve thermal contact.
\end{itemize}

\subsection{Wiper System Installation and Verification}
\begin{itemize}
    \item Place the Scotch-Brite sponge on the negative electrode.
    
    \item Insert the stainless steel rod through the designated lid slot and press it gently against the sponge.
    
    \item Secure the rod to the arm of the reciprocating actuator.
    
    \item Turn on the actuator at minimum speed and verify that:
        \begin{itemize}
            \item The stroke spans the entire electrode length
            \item The electrode surface is effectively wiped
        \end{itemize}
        
    \item Confirm that:
    \begin{itemize}
        \item The actuator is mechanically stable
        \item Alignment screws do not wobble
        \item The wiping action is gentle and does not damage the electrode
    \end{itemize}
\end{itemize}

\subsection{System Insulation and Sensor Placement}
\begin{itemize}
    \item Insulate the gel box using compressible foams, ensuring all air gaps are filled.
    
    \item Insert thermistors into:
        \begin{itemize}
            \item The gel
            \item The metal plate
        \end{itemize}
        
    \item Verify that all temperature sensors provide stable readings.
\end{itemize}

\subsection{Buffer Circulation Setup}
\begin{itemize}
    \item Connect the buffer tank to the gel box using tubing.
    
    \item Turn on the peristaltic pumps to fill the gel box until the gel is fully submerged.
    
    \item Once the desired buffer height is reached:
        \begin{itemize}
            \item Reduce the inlet pump flow rate to the target value
            \item Set the outlet pump flow rate higher than the inlet flow rate
        \end{itemize}
        
    \item Confirm that the buffer level in the gel box remains constant during circulation.
\end{itemize}

\subsection{Electrophoresis Initialization and Sample Loading}

\begin{itemize}
    \item Block the gel wells with the plugs.

    \item Connect Arduino 1 and the electrophoresis power supply to the computer.
    
    \item Open the GUI and make sure the serial communication is established.

    \item Set the voltage and timer.
    
    \item Start electrophoresis and begin monitoring gel temperature.
    
    \item Turn on the wiper.
    
    \item Allow the gel temperature to quickly rise. When the gel temperature is within 2°C of the target temperature, activate the PI temperature controller in the software.
    
    \item Once the gel temperature stabilizes at the target:
        \begin{itemize}
            \item Turn off the electrophoresis power supply
            \item Remove the desired plug
            \item Pipette samples into the gel wells
            \item Put the plug back into the well. 
        \end{itemize}
        
    \item Reset the experiment timer and restart electrophoresis to begin the run.
    
\end{itemize}

\subsection{Run Monitoring and Shutdown}
1. Monitor 
\begin{itemize}
    \item Gel temperature
    \item Buffer level
    \item Wiper operation
\end{itemize}
2. At the completion of the run:
\begin{itemize}
    \item Turn off the electrophoresis power supply
    \item Stop the wiper system
    \item Turn off pumps, heaters, and mixers
\end{itemize}

%Provide detailed instructions for the safe and proper operation of the hardware. 
%> Step-by-step operational instructions for operating the hardware. 
%> Use visual instructions as necessary. 
%> Highlight potential safety hazards.

\section{Validation and characterization}
%Demonstrate the operation of the hardware and characterize its performance over relevant critical metrics
%> Demonstrate the use of the hardware for a relevant use case. 
%> If possible, characterize performance of the hardware over operational parameters. 
%> Create a bulleted list that describes the capabilities (and limitations) of the hardware. For example consider descriptions of load, operation time, spin speed, coefficient of variation, accuracy, precision and etc. 

The performance of ARTGEL is validated by assessing its ability to maintain stable gel temperature, minimize spatial temperature gradients, and maintain constant voltage across gel during 24 hour agarose gel electrophoresis of interacting DNA origami samples. 

\subsection{Temperature stability during agarose gel electrophoresis}

\begin{figure*}[th!]
 \centering
 \includegraphics[width=1.0\linewidth]{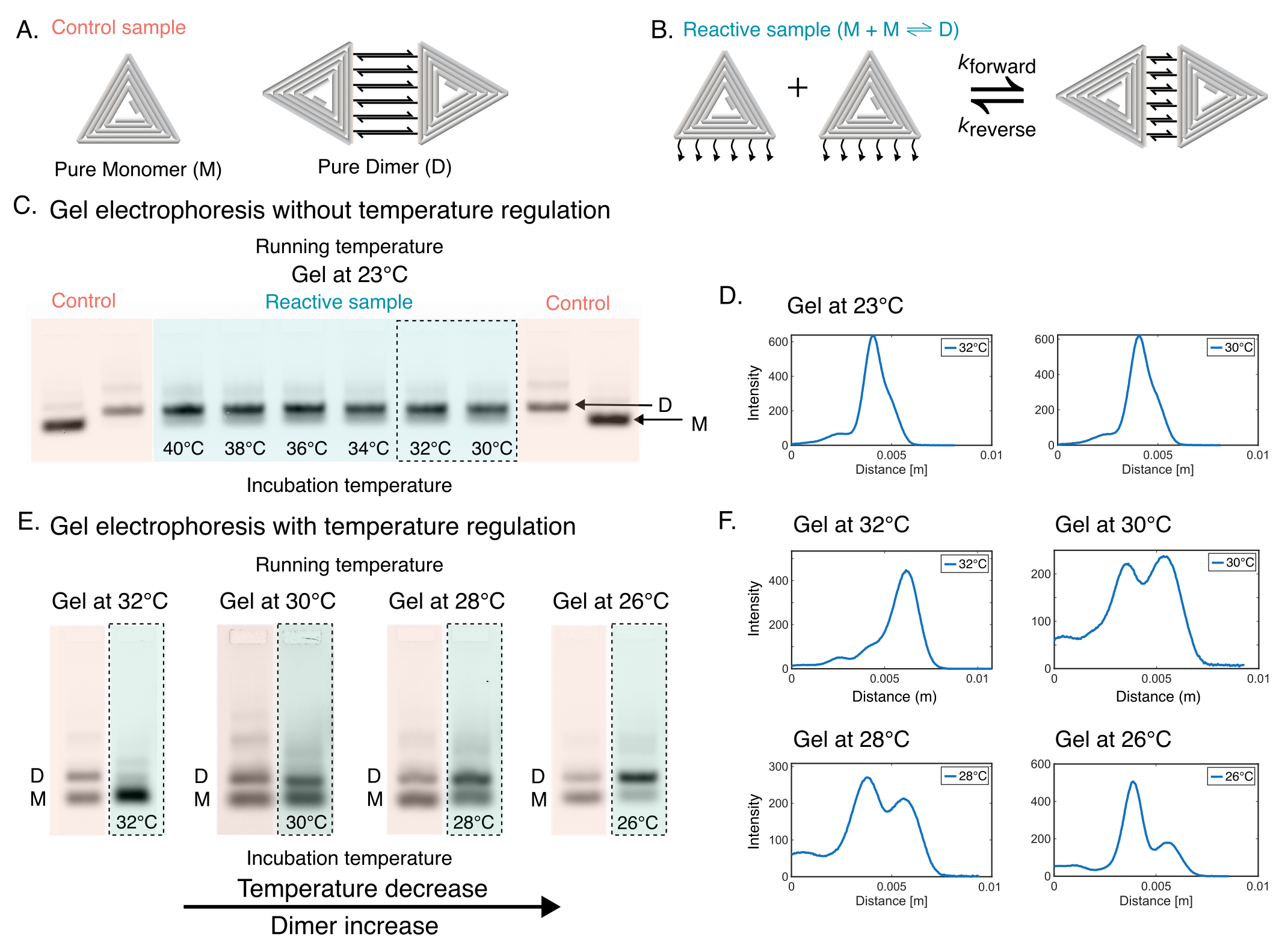}
 \caption{\textbf{Temperature dependent behavior of DNA origami building blocks in agarose gel electrophoresis.}
  A. Non-reactive DNA origami monomer  and dimer  as control samples.
  B. Reactive samples that include one side active monomer and dimer with a six-base-pairs-long hybridized domain for each overhang pair. 
  C. Agarose gel electrophoresis at ambient temperature ($23^\circ$C), without active temperature regulation. The orange and teal colored lanes contain the control and reactive samples respectively.
  Reactive samples are pre-incubated at different temperatures ranging from $40^\circ$C to $30^\circ$C. 
  D. Intensity profiles of reactive samples incubated at $30^\circ $C and $32^\circ $C and run at $23^\circ $C. The corresponding lanes in panel C are indicated with a dotted rectangular box.
  E. Agarose gel electrophoresis conducted at actively regulated constant temperatures of $32^\circ $C, $30^\circ $C, $28^\circ $C and $26^\circ $C (from right to left).
  F. Intensity profiles of the reactive samples shown in panel E}
 \label{fig:motivation}
\end{figure*}

\begin{figure*}[h!]
 \centering
 \includegraphics[width=\linewidth]{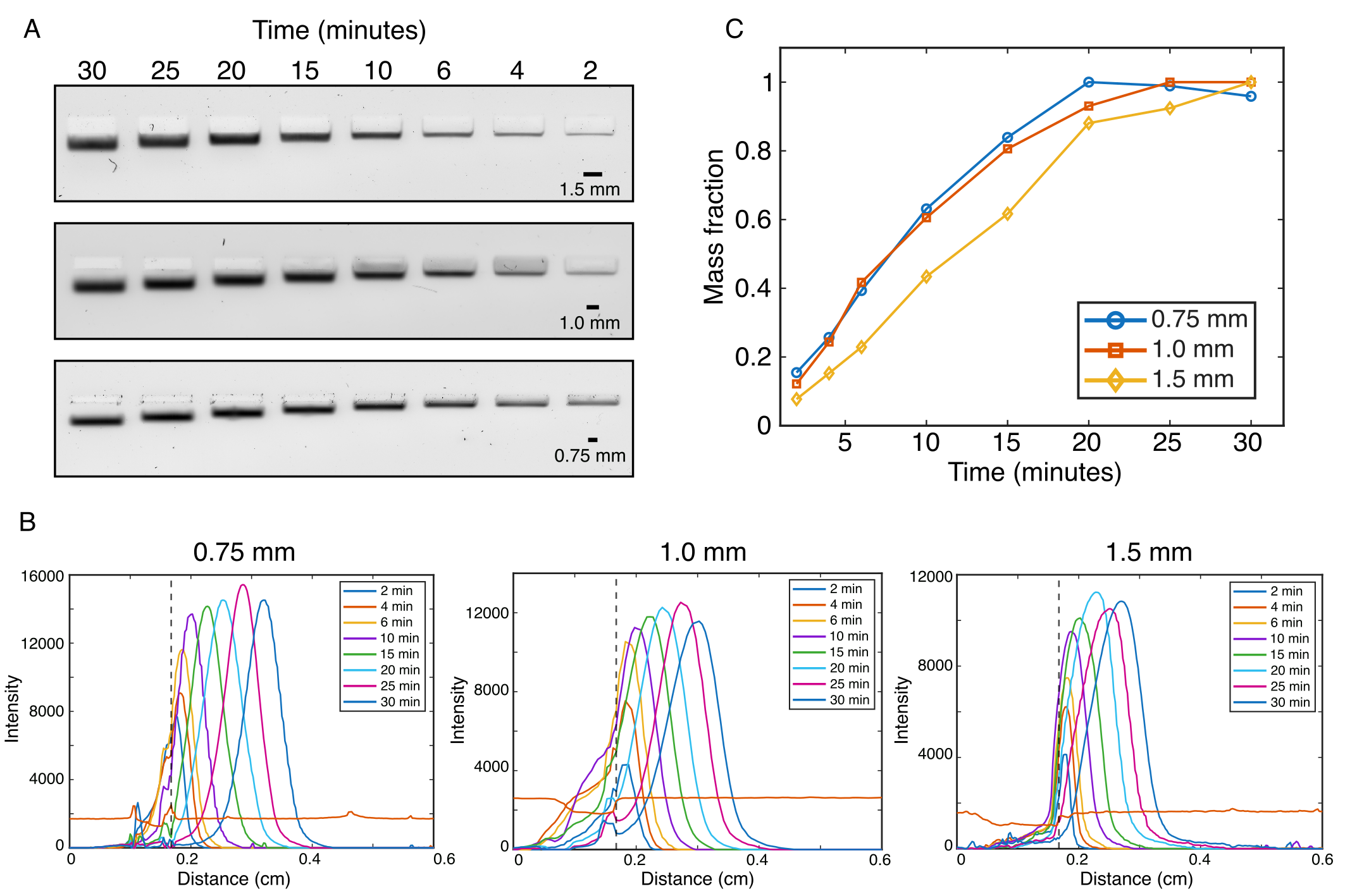}
 \caption{\textbf{Gel electrophoresis with varying well width.}
(A) Gel electrophoresis (1.5\% agarose gels, 50~V) with wells of three different widths: 1.5~mm, 1.0~mm, and 0.75~mm (top to bottom). The same volume of passivated monomer sample, 6 microliter, is loaded in each well at different time points. The total run time of sample at each well is recorded. (B)  Background-subtracted intensity profiles of the corresponding DNA bands. The distance axis refers to the vertical axis along the electrophoresis direction of DNA including the well and the band. The dotted vertical line indicates the boundary of the well, beyond which DNA has entered the agarose gel.  (C) Mass fraction as a function of run time, where the mass fraction is defined as the ratio of the area under an individual intensity profile to the maximum measured area. For this analysis, only the area beyond the dotted line in panel (B) was included. It takes about 15 minutes for 80\% of the origami in the 0.75~mm and 1.0~mm wells to enter the gel and about 20 minutes for 80\% of the origami in the 1.5~mm well to enter the gel.} 
 \label{sfig:pokt_exit}
\end{figure*}

We first examine the impact of unregulated gel temperature on temperature-sensitive DNA origami samples during agarose gel electrophoresis~\cite{hayakawa2024symmetry, saha2025modular, wei2025economical, price2025toroids}. The origami samples consisted of two controls, a ``permanent'' monomer without any designed interparticle bonds (size 50~nm) and a ``permanent'' dimer consisting of a pair of strongly bound monomers (Fig.~\ref{fig:motivation}A), and a reactive sample (Fig.~\ref{fig:motivation}B) in which the origami was capable of reversible monomer–dimer interconversion. The reactive samples were pre-incubated at temperatures ranging from $40^{\circ}$C to $30^{\circ}$C, allowed to establish distinct equilibrium states and subsequently run together in the same gel in different lanes at room temperature ($23^{\circ}$C) without actively regulated gel temperature. Despite their different initial conditions, the electrophoresis results show that all reactive samples converge to nearly identical intensity profiles, which as shown in Fig.~\ref{fig:motivation}C consists primarily of dimers. Note, that after loading, DNA origami samples require approximately 10–20 minutes to enter the gel from the well, providing sufficient time for re-equilibration (Fig.~\ref{sfig:pokt_exit}).  %Given the ambient temperature of $23^{\circ}$C, the samples most likely relax toward this temperature during migration, erasing their initial thermodynamic states and resulting in nearly indistinguishable band patterns (Fig.~\ref{fig:motivation}D).

When electrophoresis is conducted using ARTGEL under actively regulated constant gel temperatures, the initial thermodynamic states are preserved and directly resolved. DNA origami samples were incubated at temperatures ranging from $26^{\circ}$C to $32^{\circ}$C prior to electrophoresis. The reactive samples exhibit a monotonic increase in the dimer population with decreasing temperature~(Fig.~\ref{fig:motivation}E-F), in contrast to the case when samples were run at $23^{\circ}$C and the distributions were the same, independent of incubation temperature, demonstrating that ARTGEL maintains a stable thermal environment which preserves the thermodynamic states during electrophoresis.  

\subsection{Temperature regulation during electrophoresis}

Next we evaluate the temperature control performance of ARTGEL, conducting a 16-hour gel electrophoresis experiment at 50~V, with 1.5\% agarose gel. The gel and buffer reservoir are regulated at set temperatures of $30^{\circ}$C and $31^{\circ}$C, respectively. The buffer temperature is maintained $1^{\circ}$C higher than the gel to compensate for heat loss during buffer circulation.

\begin{figure*}[th!]
 \centering
 \includegraphics[width=1.0\linewidth]{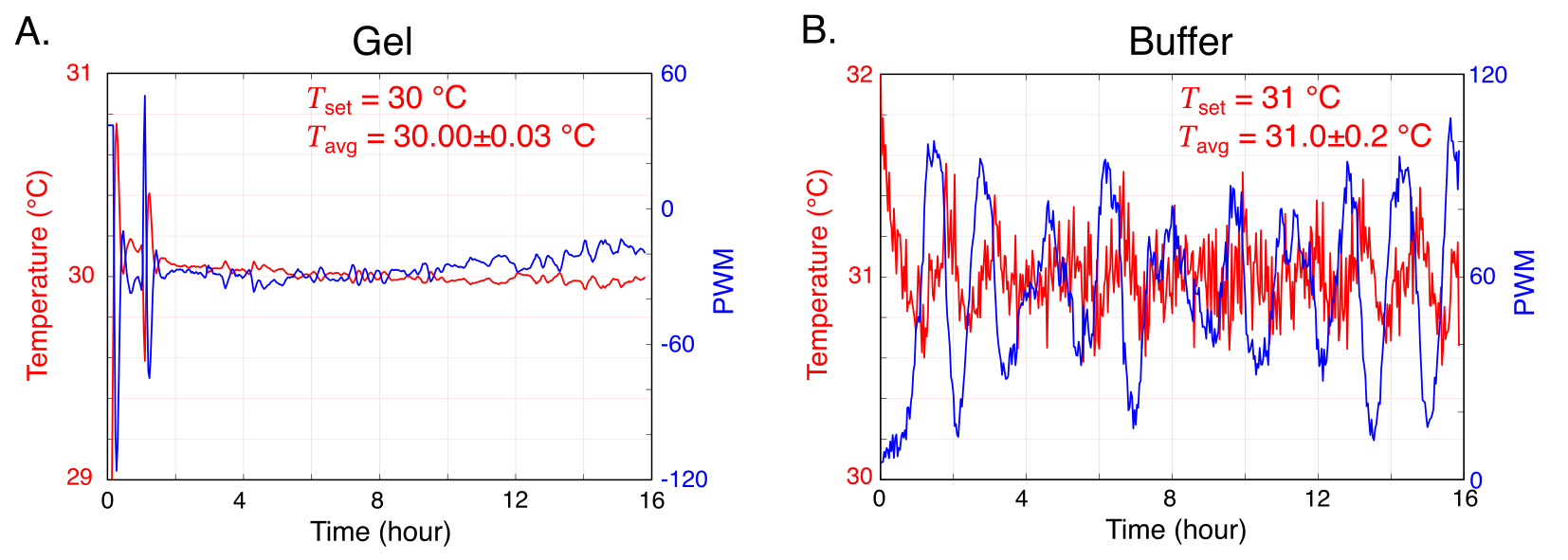}
 \caption{\textbf{Temperature regulation during agarose gel electrophoresis.}
 Agarose gel electrophoresis is conducted at 50V. The set temperatures for the gel (A) and buffer (B) are $30^\circ $C and $31^\circ $C respectively.
 Temperature (red) and corresponding PWM control signal (blue) are plotted as a function of time.}
 \label{fig:operation}
\end{figure*}

Both gel and buffer temperatures remain stable throughout the experiment. After excluding the initial 2~hours required for equilibration, the gel temperature is maintained at the set point within $\pm 0.03^{\circ}$C  (Fig.~\ref{fig:operation}A), while the buffer temperature remains within $\pm 0.2^{\circ}$C  (Fig.~\ref{fig:operation}B).

The proportional–integral (PI) control parameters are set to $k_p = 150$ and $k_i = 0.03$ for the thermoelectric module associated to the gel, and $k_p = 20$ and $k_i = 5$ for the buffer module. The buffer PWM output exhibits slow oscillations with a period of approximately one hour, arising from non-optimized controller gains (Fig \ref{fig:operation}B). However, due to the large thermal time constant of the buffer reservoir, these oscillations do not affect the buffer tank or gel temperature stability and can be further minimized through controller tuning if required.

These results demonstrate that ARTGEL provides robust and stable temperature regulation during long-duration electrophoresis experiments.

\subsection{Spatial temperature uniformity within the gel}

\begin{figure*}[th!]
 \centering
 \includegraphics[width=1.0\linewidth]{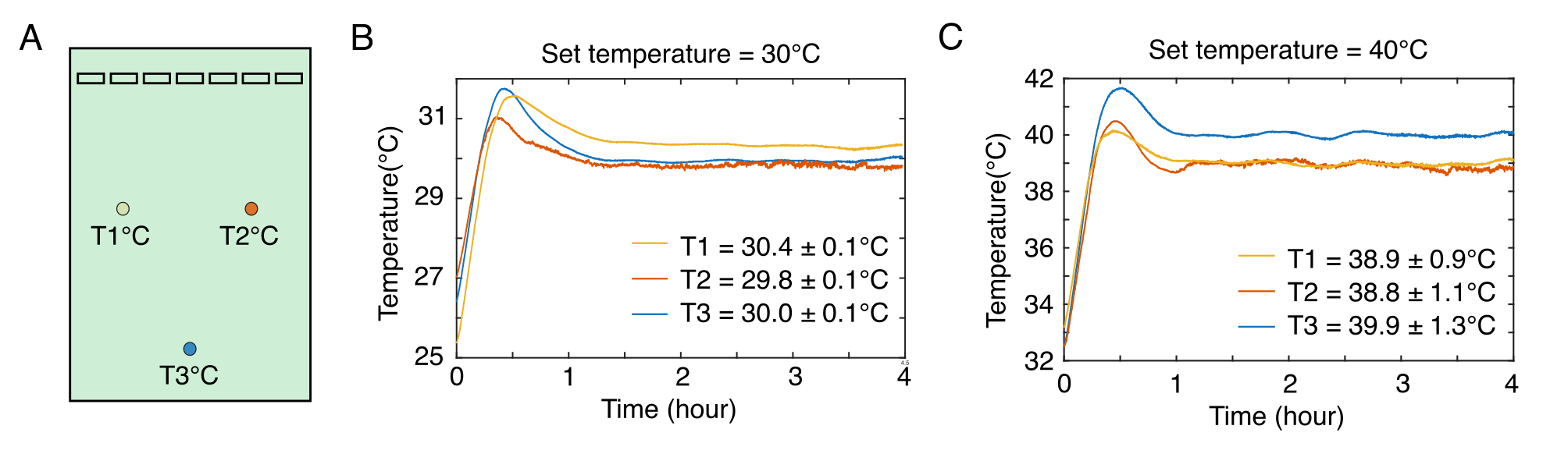}
 \caption{\textbf{Temperature distribution in the agarose gel}
 A. Three Locations in the gel where temperature have been measured simultaneously, labeled T1, T2, and T3. 
 B. Temperature as a function of time during active temperature regulation at $30^\circ $C.
 C. Temperature as a function of time during active temperature regulation at $40^\circ $C. }
 \label{fig:geltemp_dist}
\end{figure*}

Temperature uniformity across the gel is critical for preventing unintended spatial variation in electrophoretic mobility and reaction equilibria. First we calibrate all the thermistors in the temperature range from $30^\circ $C to $40^\circ $C. The calibration procedure is described in the supplement, section~\ref{SI section: calibration}. Then we measure gel temperature simultaneously at three locations during electrophoresis under active regulation at setpoints of $30^\circ $C and $40^\circ $C (Fig.~\ref{fig:geltemp_dist}A).

At a $30^\circ$C setpoint~(Fig.~\ref{fig:geltemp_dist}B), the temperature variation across the gel remains within $\pm0.4^\circ $C, while at $40^\circ$C~(Fig.~\ref{fig:geltemp_dist}C), the variation increases modestly to $\pm1.0^\circ$C. In both cases, the buffer temperature is maintained slightly above the gel temperature, which helps compensate for the heat loss during buffer circulation. These results confirm that ARTGEL provides spatially uniform temperature control suitable for quantitative electrophoresis experiments.

\subsection{Constant-voltage operation across the gel during extended runs}

To evaluate the effectiveness of automated electrode-wiping during long-duration electrophoresis, we perform 24-hour runs with and without active wiping. Under high-salt conditions (20~mM \ce{MgCl2}),  significant electrochemical deposition on the electrodes occurs, increasing their resistance. As the circuit resistance rises, the current decreases and, consequently, the voltage across the gel also drops (SI section~1). 

\begin{figure*}[th!]
 \centering
 \includegraphics[width=1.0\linewidth]{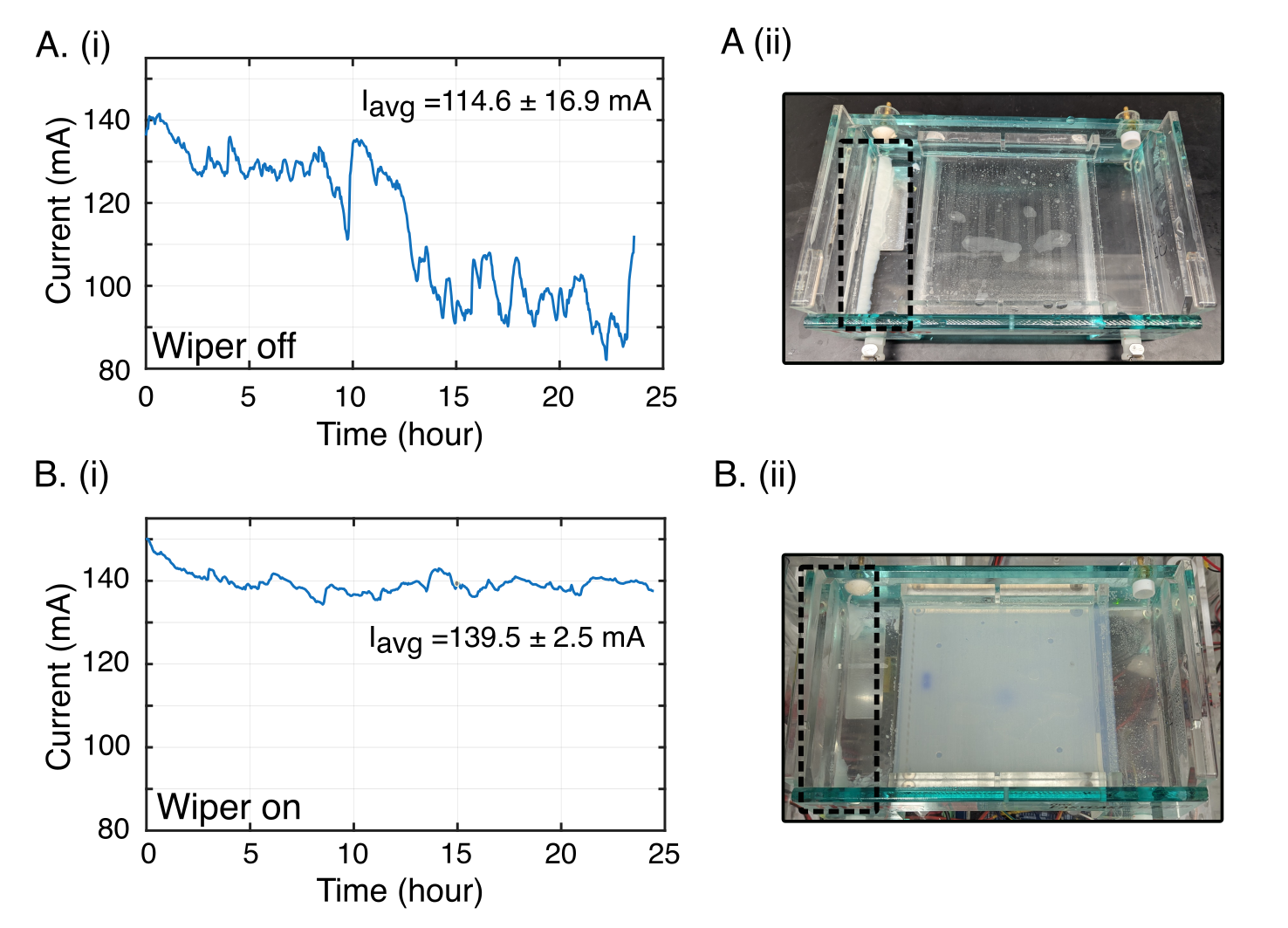}
 \caption{\textbf{Wiping performance during gel electrophoresis. }
 Current as a function of time measured in the absence of wiper (A) and during active wiping (B). Electrochemical deposition (\ce{Mg(OH)2}) on the negative electrode after electrophoresis, indicated by the dotted rectangular box.}
 \label{fig:wiping}
\end{figure*}

To quantify this effect, we monitor the electrophoresis current using the programmable electrophoresis power supply as a proxy for electrode condition and system stability. Without electrode wiping, the current decreases substantially from approximately 140~mA to 80~mA over 24 hours (Fig.~\ref{fig:wiping}A), indicating progressive electrode fouling. In contrast, with periodic automated wiping, the current remains stable at $139.5 \pm 2.5$~mA throughout the experiment (Fig.~\ref{fig:wiping}B). This stability demonstrates that the wiping module effectively mitigates salt accumulation, maintaining low electrode resistance and enabling sustained, constant-voltage operation during extended electrophoresis.

\subsection{Experiment to quantify reversible dimerization in agarose gel using ARTGEL}

\begin{figure*}[th!]
 \centering
 \includegraphics[width=1.0\linewidth]{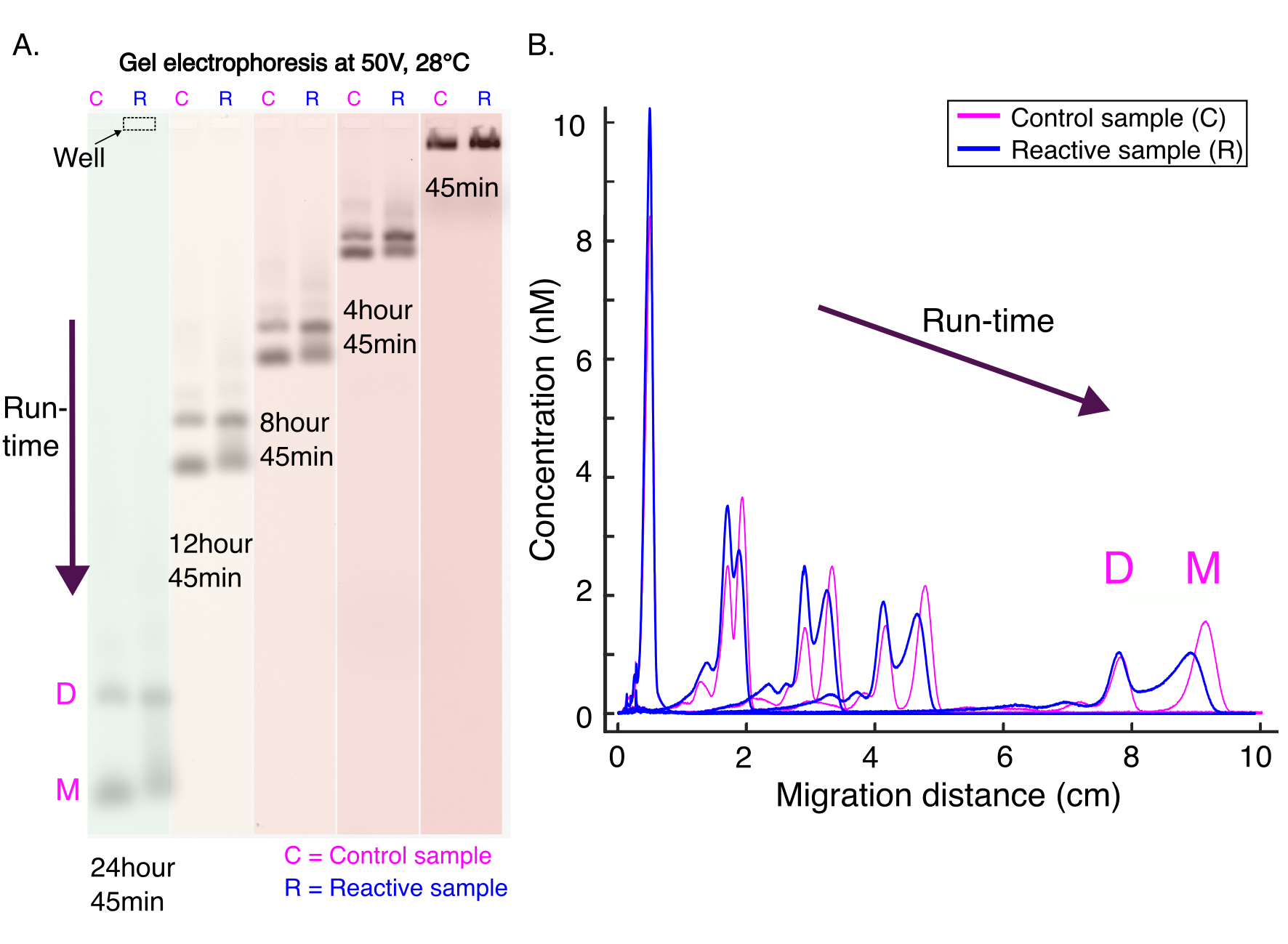}
 \caption{\textbf{Gel electrophoresis experiment to extract in-gel kinetic and thermodynamic parameters for DNA origami dimerization using ARTGEL.}
 A)  DNA origami control and reactive samples are loaded in 1.5\% agarose gel at different time points while gel temperature is maintained at $28^{\circ}$~C. The well, where the sample was loaded, is visible in the top of the image as faint, light rectangular regions directly under the labels ``C'' and ``R''. The gel was scanned using a GE Typhoon FLA 9500 Imager. 
(B) The scanned fluorescent intensity, which is proportional to the concentration as a function of position, of the control and reactive samples.}
 \label{fig:example}
\end{figure*}

Finally, we demonstrate an experiment for the extraction of in-gel binding affinity and kinetic parameters of DNA origami dimerization reaction using ARTGEL. Control samples and reactive samples are first pre-equilibrated at $28^{\circ}$C and then loaded into adjacent lanes of a 1.5\% agarose gel maintained at the same temperature (Fig.~\ref{fig:example}A). Additionally, this electrophoresis experiment is performed at the same temperature as complementary bulk light scattering experiments in the same buffer at thermodynamic equilibrium without the gel, hence the resulting in-gel parameters can be compared directly with bulk measurements to determine how confinement and migration in the porous matrix modify reversible self-assembly.

To capture the temporal evolution of the reaction, samples are loaded sequentially at multiple time points during a single electrophoresis run, resulting in five distinct datasets. The earliest loaded samples migrate for approximately 24~h~45~min, while the final set migrates for only 45~min. Throughout the experiment, both gel and buffer temperatures are actively regulated, and the electrophoresis current is continuously monitored to ensure stable operating conditions.

After electrophoresis, the gel is scanned and band intensity profiles are extracted (Fig.~\ref{fig:example}B). The control samples—consisting of permanent monomers and dimers—separate into two well-resolved peaks, reflecting their distinct electrophoretic mobilities. In contrast, the reactive samples exhibit a continuous intensity distribution between these peaks. This smearing arises from ongoing monomer–dimer interconversion during migration: monomers associate into dimers and slow down, while dimers dissociate into monomers and accelerate. The reactive dimer peak   always has present some monomer and hence on average runs a bit faster than the control dimer peak. Likewise, the reactive monomer peak always has some dimer present and on average runs a bit slower than the control monomer peak. This is clearly visible in Figs.~\ref{fig:example}A \& B by examining the position of the control and reactive bands in side-to-side comparisons at the same timepoints. Additionally, Figs.~\ref{fig:example}A \& B show that both reactive peaks become skewed and develop characteristic shoulders. The reactive monomer peak becomes skewed only on the ``slow'' side, widening towards the dimer peak and the reactive dimer peak becomes skewed on the ``fast'' side, towards the monomer peak.

To analyze the evolution of these asymmetric peak profiles across time, and to extract in-gel kinetic rates and an equilibrium parameter for the gel environment, we simultaneously fit all the concentration profiles to a set of reaction-diffusion-advection equations, details of which will be presented in a separate publication. To enhance the sensitivity of the fit it is crucial to run the experiments long enough so that the monomer and dimer peaks are separated by several times their width. To do so in this case, the electrophoresis runs need to be at least 24 hours long, justifying the need to develop an instrument with the capabilities of ARTGEL.

An additional advantage of the electrophoretic approach is that the gel resolves the distribution of species present during the reaction. In contrast to bulk-averaged methods such as static and dynamic light scattering, which report ensemble-averaged quantities, ARTGEL reveals whether the sample contains only the assumed reacting species or also additional bands arising from side products, aggregates, or intermediates. 

\section{Conclusion}

In this work, we developed ARTGEL, a temperature-regulated agarose gel electrophoresis platform that enables long-duration experiments under independently controlled thermal and electrical conditions. By maintaining a constant gel temperature and stabilizing the voltage across the gel with an automated electrode-wiping mechanism, ARTGEL creates a well-defined in-gel environment for quantitative electrophoresis of reactive systems. These capabilities allow electrophoresis to be performed at the same temperature as complementary bulk measurements while preserving the experimental control needed to interpret reaction during migration.

Using DNA origami assemblies, we show that ARTGEL preserves distinct temperature-dependent association states, provides stable gel temperatures over many hours, maintains approximate spatial temperature uniformity within the gel, and sustains constant-voltage operation for runs exceeding 24 h. Under these conditions, reactive samples develop characteristic asymmetric profiles and shoulders indicative of ongoing association and dissociation during migration. These features can be analyzed quantitatively to extract in-gel kinetic and thermodynamic parameters rather than assuming that the gel is a passive readout of the pre-equilibrated bulk solution.

ARTGEL therefore extends the role of gel electrophoresis beyond qualitative separation. It provides a practical and open-source platform for studying how confinement in porous media influences reversible biomolecular association, for comparing in-gel behavior with bulk measurements performed under matched temperature conditions, and for spatially resolving the in-gel species distributions present in the samples before loading and that arise during migration and reaction, including fragments, side products, aggregates, or additional association states. More broadly, ARTGEL makes gel electrophoresis useful as a quantitative tool for reactive transport, reversible self-assembly, and temperature-sensitive nanoscale systems in gels.

As a modular system built from widely available components and supported by openly shared hardware, firmware, and control software, ARTGEL is readily reproducible and adaptable. Although some components presently contribute substantially to the overall cost, the design is flexible and can be further standardized or simplified for other applications. Overall, ARTGEL provides a robust and accessible platform for quantitative studies of reversible association in gels and for systematic investigation of how the gel environment modifies kinetics and equilibrium.

\noindent
\textbf{Ethics statements}\\
We confirm that our work does not involve any animal or human experiments. 
\noindent

\noindent
\textbf{CRediT author statement}\\
\noindent
\textbf{Rupam Saha:} Writing - original draft, Methodology, Investigation, Conceptualization. \textbf{Seth Fraden:} Review and editing, Conceptualization, Supervision, Project administration, Funding acquisition.

\noindent
\textbf{Declaration of competing interest}\\
\noindent
The authors declare that they have no known competing interests for this work. 

\noindent
\textbf{Acknowledgements}\\
This work is supported by the Brandeis University Materials Research Science and Engineering Center, which is funded by the National Science Foundation under award number DMR-2011846. We gratefully acknowledge Francisco Mello Jr. for his mechanical and technical expertise and for providing machine-shop support. We acknowledge Wei-Shao Wei, Michael M. Norton, Hope Zhou, Amir Moore, Akshit Aggarwal, Katsu Nishiyama, Pragya Arora and Quang Tran for the helpful discussions. 

% [List here those individuals who provided help during the research (e.g., providing language help, writing assistance or proof reading the article, etc.).] Please also identify who provided financial support for the conduct of the research and/or preparation of the article and to briefly describe the role of the sponsor(s), if any, in study design; in the collection, analysis and interpretation of data; in the writing of the report; and in the decision to submit the article for publication. If the funding source(s) had no such involvement then this should be stated.}
\newpage

\textbf{Supplementary information for ``ARTGEL: A temperature-regulated electrophoresis platform for quantitative studies of reversible association in gels"}\\ 
% \textit{Please avoid acronyms and abbreviations where possible.}

%Insert Authors
\textbf{Authors}\\ \textit{Rupam Saha, Seth Fraden*}

%Insert Affiliations
\textbf{Affiliations}\\ \textit{Martin A. Fisher School of Physics, Brandeis University, Waltham, Massachusetts 02453, USA}

%Insert Contact Email
%Include institutional email address of the corresponding author
\textbf{Corresponding author’s email address}\\ \textit{fraden@brandeis.edu}

%%%%%%%%%%%%%%%%%%%%%%%%%
\section{Two resistors model for electrode fouling} \label{sec:tworesistormodel}

The electrophoresis system can be modeled as two electrical resistances connected in series: a constant gel resistance, $R_g$, and a time-dependent electrode resistance, $R_e(t)$, combining both electrodes, which increases due to salt deposition at the electrode surface. The total circuit resistance is therefore
\begin{equation}
R_{\mathrm{tot}}(t) = R_g + R_e(t).
\end{equation}

Under constant applied voltage, $V_{\mathrm{app}}$, the electrophoresis current is given by Ohm's law:
\begin{equation}
I(t) = \frac{V_{\mathrm{app}}}{R_g + R_e(t)}.
\end{equation}
As salt accumulates on the electrode, $R_e(t)$ increases, resulting in a reduction of the electrophoresis current.

The voltage drop across the gel, which determines the electric field experienced by the migrating species, is
\begin{equation}
V_g(t) = I(t)\,R_g = V_{\mathrm{app}}\,\frac{R_g}{R_g + R_e(t)}.
\end{equation}

For a gel of length $L$, the electric field within the gel is
\begin{equation}
E(t) = \frac{V_g(t)}{L}
     = \frac{V_{\mathrm{app}}}{L}\,\frac{R_g}{R_g + R_e(t)}.
\end{equation}

The electrophoretic velocity is proportional to the electric field,
\begin{equation}
v(t) = \mu\,E(t),
\end{equation}
where $\mu$ is the electrophoretic mobility. Substituting for $E(t)$ yields
\begin{equation}
v(t) = \mu\,\frac{V_{\mathrm{app}}}{L}\,\frac{R_g}{R_g + R_e(t)}.
\end{equation}

An increase in electrode resistance due to salt deposition therefore leads to a decrease in current, electric field, and electrophoretic velocity over time. Active electrode wiping suppresses the growth of $R_e(t)$, stabilizing the electric field and ensuring constant electrophoretic velocity during long-duration electrophoresis.

\section{Electrochemical Depletion of MgCl$_2$ During Electrophoresis} \label{sec:salt_depletion}

In electrophoresis experiments, sustained current flow through the buffer leads to electrochemical reactions at the electrodes that can alter the buffer composition. In the aqueous solution of \ce{MgCl2}, electrolysis reactions at the electrodes are typically:

\begin{equation} \label{eq:anode}
    Anode (+): \ce{2Cl-} \rightarrow \ce{Cl2}(g) + 2\ce{e-} \text{(Chloride ions are oxidized to chlorine gas)}
\end{equation} 
\begin{equation} \label{eq:cathode}
    Cathode (-): \ce{2H2O} + 2\ce{e-} \rightarrow \ce{H2}(g) + \ce{2OH-}(g) \text{(Water is reduced to hydrogen and hydroxide ions)}
\end{equation}

At the cathode, the produced \ce{OH-} ions react with \ce{Mg^2+} ions to form insoluble magnesium hydroxide:

\begin{equation}
\ce{Mg^2+} + \ce{2OH-} \rightarrow \ce{Mg(OH)2} 
\end{equation}

Our electrophoresis gel box contains approximately 0.8~L of 20~mM \ce{MgCl2} solution including both the gel and buffer. During electrophoresis, \ce{MgCl2} steadily depletes from the system. We calculate the rate of salt depletion below.%Quantifying this depletion is essential for understanding changes in ionic strength and maintaining structural stability and experimental reproducibility. Below we calculate the rate of \ce{MgCl2} depletion from our system.

% \subsection{Charge Transfer and Electron Flux}
The total electric charge passed through the process is given by:
\begin{equation}
Q = I \cdot t
\end{equation}
where $I$ is the applied current (A) and $t$ is the duration of electrophoresis (s). For an average current of $I = 0.13$~A applied over $t = 3600$~s (1~hour), the total charge is:
\begin{equation}
Q = 0.13 \times 3600 = 468 \ \mathrm{C}
\end{equation}

Using Faraday’s constant ($F = 96485$ C/mol), the number of moles of electrons transferred is:
\begin{equation}
n_e = \frac{Q}{F} = \frac{468}{96485} \approx 4.85 \times 10^{-3} \ \mathrm{mol}
\end{equation}

% \subsection{Electrode Reactions and Stoichiometry}
Since each MgCl$_2$ unit provides two chloride ions, the number of moles of \ce{MgCl2} depleted is:

\begin{equation}
n\ce{MgCl2} = \frac{n_e}{2} = \frac{4.85 \times 10^{-3}}{2} \approx 2.43 \times 10^{-3} \ \mathrm{mol}
\end{equation}

% Substituting the calculated electron flux:
% \begin{equation}
% n_{\mathrm{MgCl_2}} = \frac{4.85 \times 10^{-3}}{2} \approx 2.43 \times 10^{-3} \ \mathrm{mol}
% \end{equation}

% \subsection{Mass Loss and Concentration Change}

% The molar mass of \ce{MgCl2} is approximately 95.2 g/mol. The corresponding mass loss over 1 hour is:
% \begin{equation}
% m = n{\mathrm{MgCl_2}} \times 95.2 = 2.43 \times 10^{-3} \times 95.2 \approx 0.23 \ \mathrm{g}
% \end{equation}

The Thermo gel box holds a buffer volume of $V = 0.8$ L. With an initial \ce{MgCl2} concentration of 20 mM, the initial amount of \ce{MgCl2} is:
\begin{equation}
n_0 = 0.020 \times 0.8 = 0.016 \ \mathrm{mol}
\end{equation}

Thus, the fractional depletion per hour is:
\begin{equation}
\frac{n_{\mathrm{MgCl_2}}}{n_0} \approx \frac{2.43 \times 10^{-3}}{0.016} \approx 0.15
\end{equation}

indicating that approximately 15\% of the \ce{MgCl2} is depleted per hour under these conditions. At this rate of depletion, no \ce{Mg^2+} would be left after a 24 hour run.  To reduce the percentage change in \ce{Mg^2+} concentration over 24 hours, we circulate 8L of heated buffer through the electrophoresis chamber and we replace the 8L of buffer with fresh buffer after 12 hours.

% \subsection{Implications for Long-Duration Operation}

%This analysis demonstrates that electrochemical reactions can significantly deplete the amount of \ce{MgCl2} over experimentally relevant timescales. For extended runs (e.g., 24 hours), gradual depletion of \ce{MgCl2} may lead to substantial reductions in ionic strength and changes in electrophoretic mobility. Therefore, maintaining constant ionic conditions in long-duration electrophoresis requires either buffer replenishment, or active recirculation mechanisms to mitigate electrochemical depletion effects.
\newpage
%%%%%%%%%%%%%%%%%%%%%%%%%

\section{PI Control and Controller Tuning}\label{sec:PItuning}

We regulate the gel temperature using a proportional–integral (PI) feedback control loop that adjusts the applied pulse-width modulation (PWM) signal in response to real-time temperature measurements. This closed-loop system compensates for Joule heating generated during electrophoresis and maintains the gel at a prescribed target temperature.

The control law is implemented as
\begin{equation}
u(t) = K_p\, e(t) + K_i \int_{0}^{t} e(\tau)\, d\tau,
\end{equation}
where the error is defined as
\begin{equation}
e(t) = T_{\mathrm{set}} - T(t).
\end{equation}
Here, $u(t)$ represents the control signal, $K_p$ and $K_i$ are the proportional and integral gains, $T_{\mathrm{set}}$ is the target temperature, and $T(t)$ is the measured gel temperature.

The control signal $u(t)$ is mapped to the thermoelectric actuator through a pulse-width modulated (PWM) signal. The magnitude of $u$ determines the PWM duty cycle (ranging from 0 to 255), which sets the fraction of time that the supply voltage is applied to the thermoelectric module via an H-bridge. The sign of $u$ determines the polarity of the applied voltage, enabling both heating and cooling:
\begin{equation}
\mathrm{PWM}_k = \min\left( \max\left( |u_k|, 0 \right), 255 \right),
\end{equation}
\begin{equation}
\mathrm{Heat/Cool} = \mathrm{sgn}(u_k).
\end{equation}

Physically, the PWM duty cycle controls the average electrical power delivered to the thermoelectric module. A higher duty cycle increases the fraction of time the module is energized, thereby increasing the heat flux into or out of the gel. Due to the large thermal mass of the gel and buffer, the system responds to the time-averaged power rather than the high-frequency switching, allowing PWM to act as an effective continuous control input.

\subsection{Experimental setup}

We prepared 1.5\% agarose gel and running buffer of 20~mM \ce{MgCl2}. We prepared a total of 6~L of buffer and heated it to $32~^{\circ}$C. After the buffer temperature stabilized at the setpoint, buffer circulation between the external reservoir and the electrophoresis chamber was initiated at a flow rate of 30~mL/min with the insulation enabled. An electrophoresis voltage of 50~V was then applied, and the gel temperature was continuously monitored. Once the gel temperature reached a steady-state plateau, the following experiments were performed. Throughout all subsequent experiments, buffer circulation, insulation, and the applied electrophoresis voltage were maintained continuously.

\begin{figure*}[h!]
 \centering
 \includegraphics[width=\linewidth]{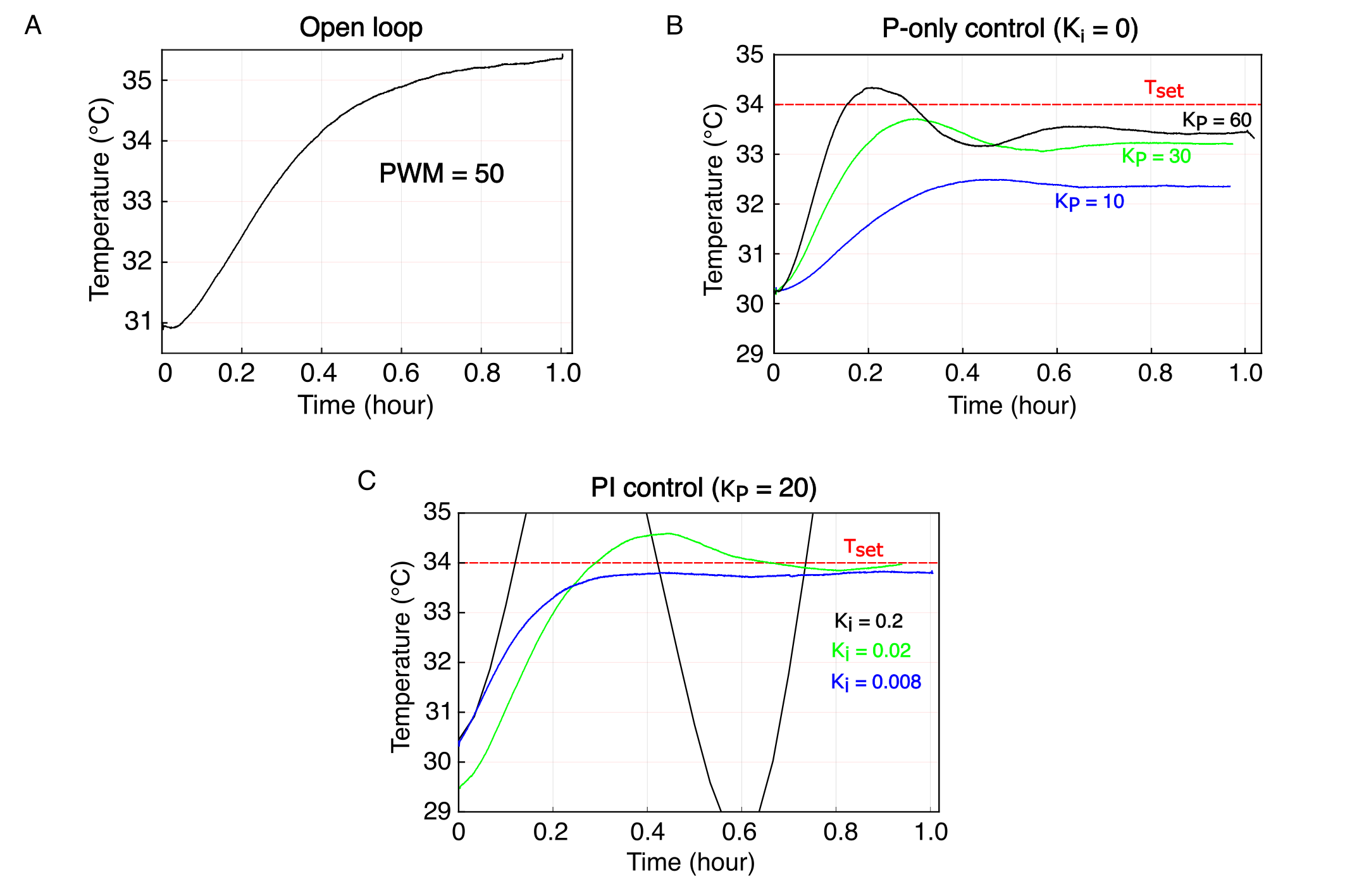}
 \caption{\textbf{Parameter sweep for proportional-integral control of the gel.}
 (A) Manual or open loop control.  
(B) $K_P$ sweep
(C) ki sweep when $K_{p}$ = 20}
 \label{sfig:PItuning}
\end{figure*}

\subsection{Open-loop thermal response}

To characterize the intrinsic thermal dynamics of the system, we first performed an open-loop experiment (SI Fig.~\ref{sfig:PItuning}A). A constant PWM value of 50 was applied for one hour, resulting in an increase in gel temperature from 31$^{\circ}$C to 35$^{\circ}$C, after which the system approached a steady-state plateau. This slow response indicates a large thermal time constant on the order of tens of minutes, reflecting the combined heat capacity of the gel and surrounding buffer.

\subsection{Proportional control (K\textsubscript{p} sweep)}

We next evaluated proportional (P) control by setting $K_i = 0$ and varying $K_p$ (SI Fig.~\ref{sfig:PItuning}B). In this regime, the control signal is directly proportional to the instantaneous temperature error. While this improves response speed, it inherently produces a steady-state offset (droop), since a nonzero error is required to sustain the control action.

At low gain ($K_p = 10$), the system approaches the setpoint (34$^{\circ}$C) slowly and exhibits a steady-state error of 1.8$^{\circ}$C. Increasing the gain to $K_p = 30$ reduces the steady-state error to 0.8$^{\circ}$C and accelerates convergence. At high gain ($K_p = 60$), the system responds rapidly but exhibits oscillatory behavior due to overcorrection in the presence of thermal delay, settling with a residual error of approximately 0.6$^{\circ}$C. These results demonstrate the trade-off between responsiveness and stability, with $K_p \approx 20$–30 providing a practical compromise.

\subsection{Proportional–integral control (K\textsubscript{i} sweep)}

To eliminate steady-state error, we introduced an integral term, which accumulates the error over time and drives the system toward the exact setpoint. However, the integral gain must be carefully tuned for slow thermal systems.

We fixed $K_p = 20$ and varied $K_i$ (SI Fig.~\ref{sfig:PItuning}C). At high integral gain ($K_i = 0.2$), the system exhibits large-amplitude, long-period oscillations (approximately 30 min) due to rapid accumulation of the integral term, leading to overshoot and slow recovery. At moderate gain ($K_i = 0.02$), the system achieves stable convergence to the setpoint with minimal oscillation. At low gain ($K_i = 0.008$), convergence is slower and small residual errors persist over extended times.

\subsection{Optimal control parameters}

Together, these results indicate that stable and efficient temperature regulation is achieved with $K_p \approx 20$ and $K_i \approx 0.02$. Under these conditions, the system reaches a 4$^{\circ}$C temperature increase within approximately 30 minutes while maintaining minimal oscillation and negligible steady-state error.

\clearpage

\section{Use of plugs}\label{sec:plug}

\begin{figure*}[th!]
 \centering
 \includegraphics[width=0.7\linewidth]{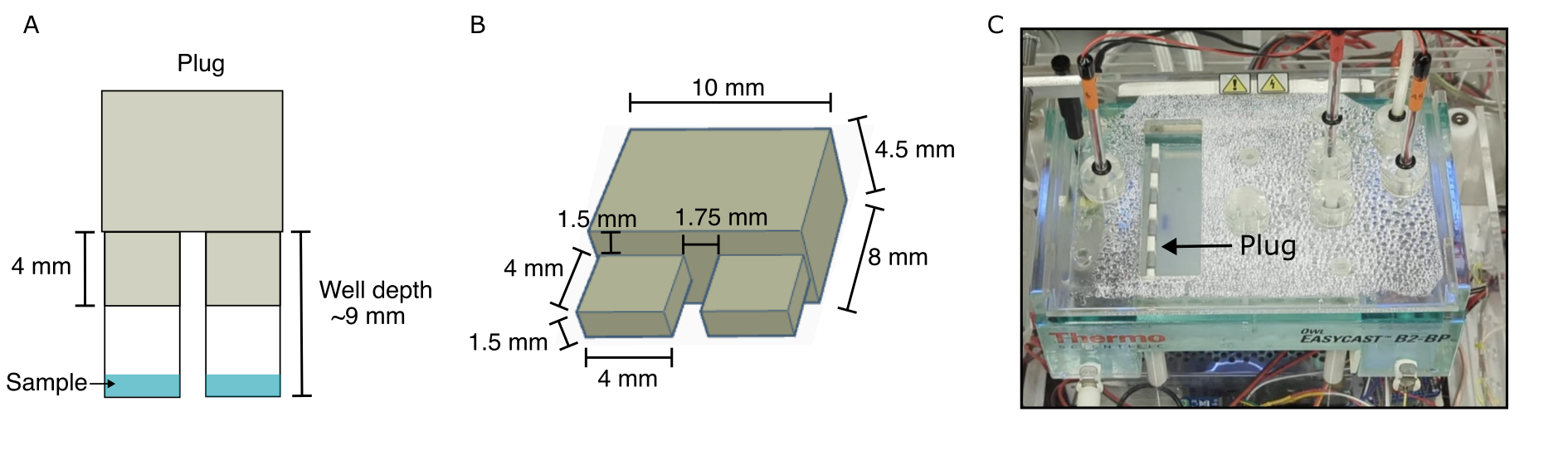}
 \caption{\textbf{3D-printed Plug to block the gel wells}
(A)A plug with two teeth placed in wells. (B) 3D design of a plug with measurements. (C) An image of a gel where the wells are blocked with plugs.}
 \label{sfig:plug1}
\end{figure*}

 We designed and 3D printed well plugs  for gels with 1.5 mm thick wells. The wells had dimensions $l\times w \times h = 4 \times 1.5 \times 9$ mm$^3$ and the plug  had dimensions $l\times w \times h = 4 \times 1.5 \times 4$ mm$^3$ (Figure~\ref{sfig:plug1}). Each plug features two teeth, allowing it to anchor securely across adjacent wells and improve stability during operation.

\begin{figure*}[h!]
 \centering
 \includegraphics[width=0.45\linewidth]{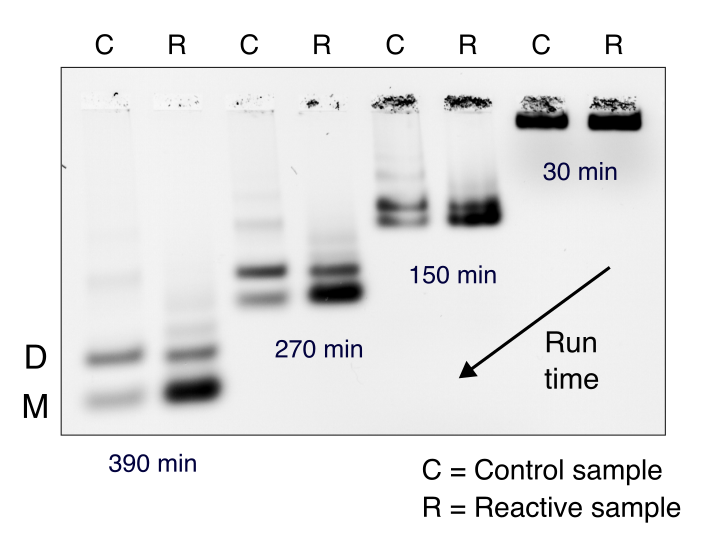}
 \caption{\textbf{Gel electrophoresis with open wells (without  plugs).}
A laser-scanned micrograph of a gel loaded with control and reactive DNA origami samples.}
 \label{sfig:plug2}
\end{figure*}

Figure~\ref{sfig:plug2} shows a representative gel in which control and reactive samples are loaded into adjacent lanes at sequential time points, in which the lanes on the left are loaded at the beginning of the experiment and those on the right are loaded after 6 hours of electrophoresis. During the initial loading step, the wells show no fluorescence, but over time fluorescence from the DNA origami increases in the wells. Our hypothesis is that the fluorescence comes from origami complexed with \ce{Mg(OH)2} precipitants. No precipitant forms in the beginning of the experiment. However, as the experiment progresses, the electrode wipers continuously remove \ce{Mg(OH)2} from the negative electrode. Some of this material subsequently redeposits into the wells. We hypothesize that the DNA origami loaded late in the run becomes adsorbed to, or physically trapped in this hydrated hydroxide deposit. As a result, the concentration of DNA origami that moves through the gel at late times will be less than DNA origami injected into the gel at early times.  

Overall, the use of plugs helps maintain the concentration of origami in the gel during the 24 hour period of gel electrophoresis.

\clearpage

\section{Calibration of thermistor} \label{SI section: calibration}
Prior to installation, all thermistors were calibrated in a stirred water bath together with a reference digital thermocouple over a temperature range $20^{\circ}$ C - $45^{\circ}$ C. For each thermistor, the measured divider voltage was first converted to resistance and then to a raw temperature value using Steinhart–Hart equation.  

\begin{equation}
\frac{1}{T} = \frac{1}{T_{0}} + \frac{1}{\beta}\ln\left(\frac{R}{R_{25}}\right)
\end{equation}

 Here $\beta = 4540~\mathrm{K}$ and $R_{25} = 100~\si{\kilo\ohm}$. 
 
 The corresponding equilibrium temperatures measured by the reference thermocouple were recorded simultaneously. A linear regression was then performed for each thermistor according to
\begin{equation}
T_m = mT_{\mathrm{ref}} + c
\end{equation}
where $T_m$ is the raw thermistor-reported temperature, $T_{\mathrm{ref}}$ is the thermocouple temperature, and $m$ and $c$ are the sensor-specific slope and intercept, respectively. During subsequent instrument operation, each thermistor reading was corrected using the inverse relation
\begin{equation}
T_r = \frac{T_m - c}{m}
\end{equation}
where $T_r$ is the calibrated reported temperature. This procedure compensates for sensor-to-sensor variation while retaining a common nominal thermistor model for all sensing channels. 

\begin{figure*}[h!]
 \centering
 \includegraphics[width=\linewidth]{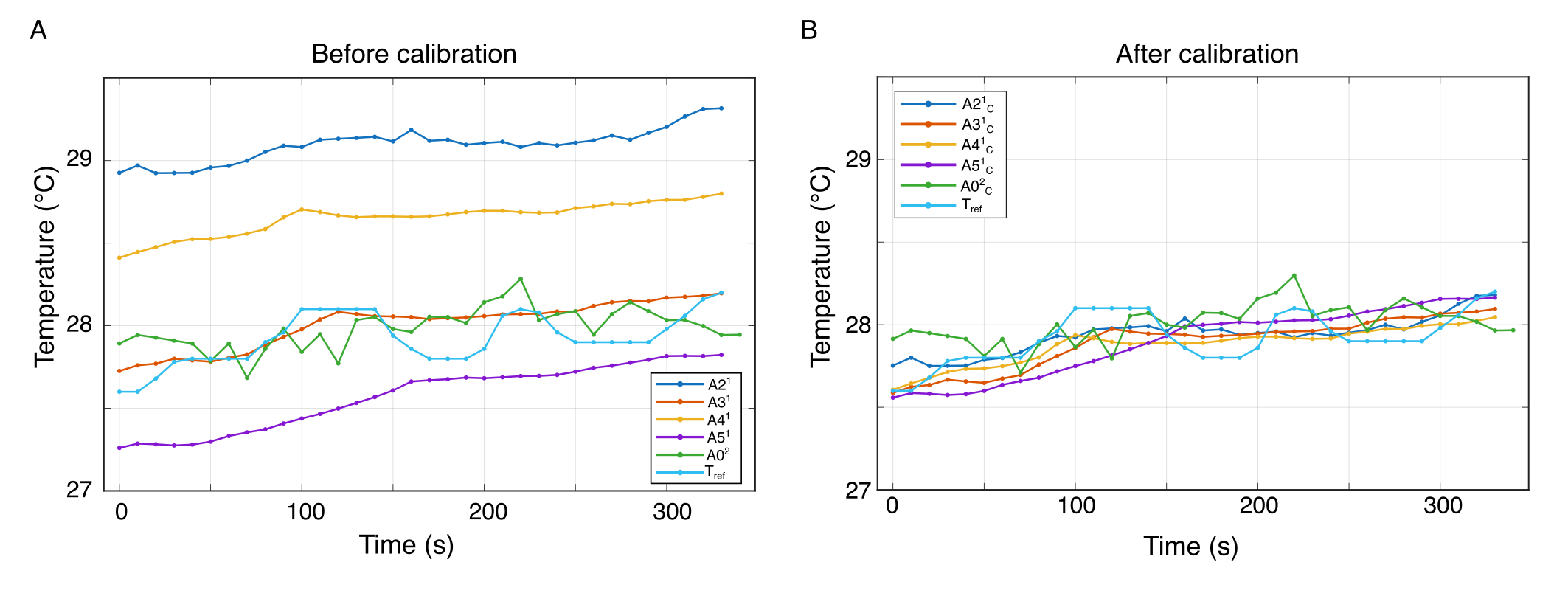}
 \caption{\textbf{Calibration of thermistors using a reference thermocouple.} 
(A) Temperature versus time measured by different thermistors (see Fig.~4 for thermistor labels) and the reference thermocouple ($T_{\mathrm{ref}}$) before calibration. (B) Temperature versus time for the calibrated thermistors, replotted together with the reference thermocouple ($T_{\mathrm{ref}}$).
} 
 \label{sfig:calibration}
\end{figure*}

\clearpage

\section{Measurement of the time required for DNA to exit the gel well}
\label{SI section: exit_pokt}

 The time required for DNA origami monomers to exit the sample well and enter the agarose gel was experimentally evaluated using gel combs with three different well widths, $W =$~0.75~mm, 1.0~mm and 1.5~mm. The sample wells had dimensions of 4~mm (length) $\times$ $W$~mm (width) $\times$ 9~mm (depth), corresponding to a total well volumes of 27, 36 and 54~\si{\micro\liter}, respectively. All experiments were performed using 1.5\% agarose gels in running buffer containing 20~mM \ce{MgCl2}, under identical electrophoresis conditions at an applied voltage of 50~V. For each experiment, 6~\si{\micro\liter} of sample was loaded into the well, consisting of 5~\si{\micro\liter} of DNA origami sample mixed with 1~\si{\micro\liter} of loading dye. 

Following gel preparation, identical monomer samples were loaded sequentially from left to right across the gel. This staggered loading procedure allowed different effective electrophoresis times within a single run, such that the first loaded sample migrated for 30~min, whereas the final sample migrated for only 2~min before the run was terminated.

At the completion of electrophoresis, each sample well was thoroughly rinsed to remove any residual DNA remaining inside the well. The gel was then imaged, and intensity profiles extracted from the scanned image after background subtraction. For each lane, the area under the migrated DNA peak was determined. The largest peak area measured in the experiment was taken as the reference corresponding to complete sample release. A mass fraction was then defined as the ratio of the peak area for a given lane to this maximum value, representing the fraction of DNA that had exited the well.

The results indicate that approximately 20~min were required for nearly complete release of the DNA sample from the well under the tested conditions. Initially, it was hypothesized that narrower wells would shorten the exit time by reducing the distance required for DNA transport within the well. However, only a weak dependence on well width was observed. This suggests that the dominant transport resistance arises at the well--gel interface rather than within the well itself. Our physical picture is that DNA molecules are transported rapidly within the liquid-filled well and accumulate transiently at the gel boundary, while the slower migration into the agarose matrix limits the overall entry rate. Consequently, the release kinetics are largely insensitive to well width over the tested range. These observations further suggest that the gel-entry time is more strongly governed by parameters such as the applied voltage and agarose concentration, than well width.

\clearpage

\section{Supplementary movie} \label{SI:movie}

\noindent\textbf{Operation of the ARTGEL instrument.} 
\href{https://brandeis.box.com/shared/static/z4872mxgwyy0lwl6kj05fqmep1rhvugh.mp4}{Movie link}
Operation of the ARTGEL instrument during electrophoresis (shown without insulation). Buffer circulation, active wiping, and real-time monitoring of temperature and current are demonstrated during device operation.
\vspace{45mm}

\begin{figure*}[h!]
 \centering
 \includegraphics[width=0.8\linewidth]{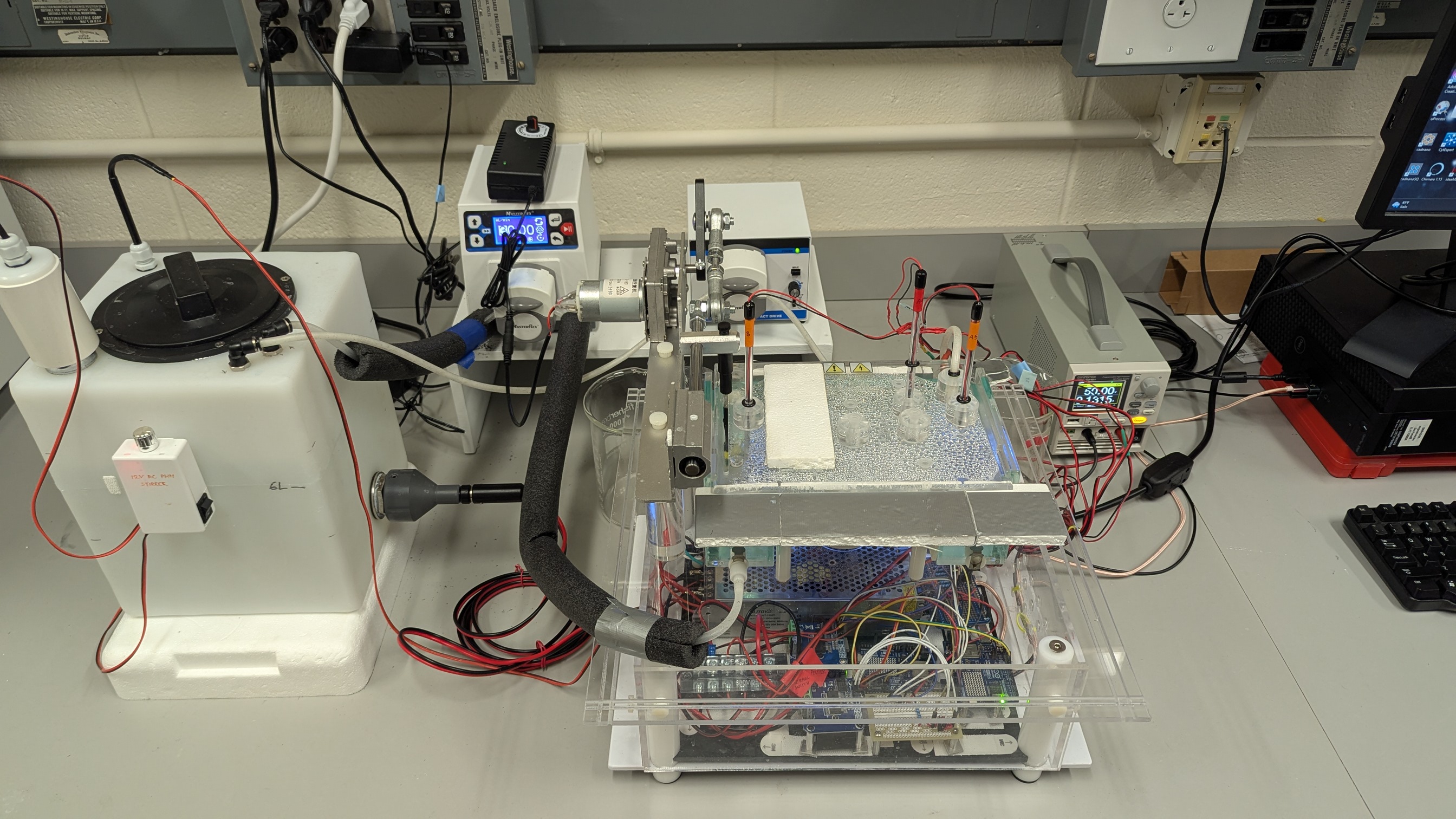}
 \caption{\textbf{ARTGEL instrument with insulated electrophoresis system.}
} 
 \label{sfig:insulation}
\end{figure*}

\bibliographystyle{unsrtnat}
\bibliography{references}

\end{document}